%% file: main.tex
\documentclass{article} 
\usepackage[utf8]{inputenc}
\usepackage[margin=1in]{geometry}
\usepackage{xr-hyper}
\usepackage[unicode=true,
bookmarks=false,
breaklinks=false,pdfborder={0 0 0},colorlinks=false]
{hyperref}

\usepackage{bibentry}
\renewcommand{\shortstack}[1]{\begin{tabular}{c}#1\end{tabular}}
\usepackage{csquotes}

\def\clap#1{\hbox to 0pt{\hss#1\hss}}
\usepackage{lmodern,textcomp}
\usepackage{bbm,url}

\author{Cannon Cloud\thanks{
		\noindent Cannon Cloud: Goethe University Frankfurt, \href{mailto:cloud@econ.uni-frankfurt.de}{cloud@econ.uni-frankfurt.de}}  \quad and \quad  
		Simon He\ss\thanks{Simon Heß: University of Vienna, \href{mailto:simon.hess@univie.ac.at}{simon.hess@univie.ac.at}}  \quad and \quad 
		Johannes Kasinger\thanks{Johannes Kasinger: Goethe University Frankfurt and Leibniz Institute for Financial Research SAFE, \href{mailto:kasinger@safe.uni-frankfurt.de}{kasinger@safe-frankfurt.de}}
		}
		
\usepackage[style=authoryear,uniquename=false,maxcitenames=2,maxbibnames=99,uniquelist=false,giveninits=true,natbib=true,citestyle=authoryear-comp]{biblatex}
\renewbibmacro{in:}{}
\renewbibmacro*{volume+number+eid}{\printfield{volume}\setunit*{\addnbspace}\printfield{number}\setunit{\addcomma\space}\printfield{eid}}
\DeclareFieldFormat[article]{number}{\mkbibparens{#1}}
\DeclareNameAlias{author}{family-given}
\usepackage{xurl,todonotes} 
\usepackage{graphicx,xcolor}
\usepackage[T1]{fontenc}
\usepackage{amsmath,amsthm,amssymb}
\usepackage{booktabs,adjustbox,threeparttable,placeins}
\usepackage{verbatim}
\usepackage{cleveref}
\newif\ifinappendix
\inappendixfalse
\crefname{appsec}{appendix}{appendices}
\Crefname{appsec}{Appendix}{Appendices}
\crefname{apptab}{\protect{\ifinappendix\else appendix \fi}table}{\protect{\ifinappendix\else appendix \fi}tables}
\Crefname{apptab}{\protect{\ifinappendix Table\else Appendix table\fi}}{\protect{\ifinappendix Tables\else Appendix tables\fi}}
\crefname{appfig}{\protect{\ifinappendix\else appendix \fi}fig.}{\protect{\ifinappendix\else appendix \fi}figures}
\Crefname{appfig}{\protect{\ifinappendix Fig.\else Appendix fig.\fi}}{\protect{\ifinappendix\else Appendix figures\fi}}
\title{Do shared e-scooter services cause traffic accidents?\\Evidence from six European countries\thanks{
		\noindent We thank Kirill Borusyak, Fabrizio Colella, Markus Eyting, Nicolas Koch, Patrick Schmidt, and Matthias Schündeln for constructive comments and suggestions. We further thank the companies Tier and Voi for providing us data on launch months in cities. Map data copyrighted OpenStreetMap contributors and available from \href{https://www.openstreetmap.org}{www.openstreetmap.org}. JK gratefully acknowledges financial support from the Leibniz Institute for Financial Research SAFE. SH thanks the Joachim Herz Foundation for  financial support.}}
\date{\today}
\renewcommand{\cite}[1]{\cdcdc}
\begin{document}
\begin{refsection}
\maketitle
\begin{abstract}
We estimate the causal effect of shared e-scooter services on traffic accidents by exploiting variation in availability of e-scooter services, induced by the staggered rollout across 93 cities in six countries. Police-reported accidents in the average month increased by around 8.2\% after shared e-scooters were introduced. For cities with limited cycling infrastructure and where mobility relies heavily on cars, estimated effects are largest. In contrast, no effects are detectable in cities with high bike-lane density. This heterogeneity suggests that public policy can play a crucial role in mitigating accidents related to e-scooters and, more generally, to changes in urban mobility.
\end{abstract}

\section{Introduction}
\input{Introduction}

\section{Data and empirical strategy}\label{sec:mm}
\input{MaterialsMethods}

\section{Results}\label{sec:results}
\input{Results}

\section{Discussion}\label{sec:discussion}
\input{Discussion}

\printbibliography\clearpage

\end{refsection}

\begin{refsection}\appendix\inappendixtrue\crefalias{section}{appsec}\crefalias{table}{apptab}\crefalias{figure}{appfig}\crefalias{subsection}{appsec}\crefalias{subsubsection}{appsec}\renewcommand{\thesection}{\Alph{section}}\renewcommand{\thesubsection}{\Alph{section}.\arabic{subsection}}\renewcommand{\thesubsubsection}{\Alph{section}.\arabic{subsection}.\arabic{subsubsection}}\renewcommand{\thefootnote}{\arabic{footnote}}\setcounter{page}{1}\part*{Supplementary appendix}\section{Additional analyses}\FloatBarrier\subsection{Two-way fixed-effects}\label{app-robustness}\FloatBarrier\input{app-robustness}\subsection{Treatment timing and instrumental variable analyses}\label{section:app-endogeneity}\input{app-endogenity}\subsection{Additional tables and figures}\FloatBarrier\label{app-more-tables}\input{app-more-tables}\clearpage\subsection{Synthetic difference-in-differences \citep{arkhangelsky2021synthetic}}\label{app-synth-did}

\input{app-synth-did}\section{Data description}\FloatBarrier\input{app-data}\section*{Appendix references}\printbibliography
\end{refsection}
\end{document}

%% file: Introduction.tex
In the EU alone, over 150,000 people are killed or seriously injured in traffic accidents every year.
The \citet{oecd2018} estimates that social costs of traffic accidents exceed 3\% of the EU's GDP. In recent years, e-scooters emerged as a prominent mode of transportation in cities worldwide. Between 2018 and mid-2022, more than \$5bn was invested in companies providing shared e-scooter services \citep{McKinsey2022}. Despite their increased availability, the role of shared e-scooters in future urban mobility ecosystems remains a highly divisive discussion topic.
Proponents argue that shared e-scooters can ease issues related to motorized traffic, such as air pollution, noise, and congestion \citep{shaheen2019shared, gossling2020integrating, abduljabbar2021role}. Opponents raise concerns about sustainability, safety, and crowded sidewalks \citep{hollingsworth2019scooters, james2019pedestrians, sanders2020scoot}.

One central point of contention is the social cost induced by shared e-scooter services through traffic accidents. Thus far, the public and academic discussion has mostly focused on the large relative change in injuries from e-scooter-related accidents over time \citep{choron2019integration}. This increases occurs by design because e-scooters were virtually non-existent before 2018.
Additionally, this approach is lacking because substitution between modes of traffic and other indirect effects are not accounted in by these raw numbers, as detailed below.

To address this scarcity of evidence, this article studies medium-run effects of the introduction of shared e-scooters on road safety in six European countries. We identify the effect on urban accidents using quasi-experimental variation in the availability of shared e-scooters across cities and time. 
In our analyses, treatment start is defined as the city-specific launch date of the first shared e-scooter service.
Once the technology and capital for dockless e-scooters became widely available and national regulation allowed their use, e-scooter firms rapidly expanded into large cities across our sample countries, Austria, Finland, Germany, Norway, Sweden, and Switzerland.
We assume that the city-specific treatment timing is exogenous to accidents conditional on city fixed effects and time fixed effects since the rollout of e-scooter services is mainly determined by national regulation and business constraints, time-invariant city characteristics, as well as seasonal variation. In more technical terms, \citet{ghanem2022selection} show that a sufficient condition for the parallel trends assumption to hold are that the selection mechanism is independent from city-time-varying unobservables and that city-time-varying unobservables that affect accidents have a constant mean over time conditional on city-level time-invariant unobservables.
This allows us to estimate effects in a difference-in-differences model, using the monthly data on traffic accidents in cities before and after the arrival of e-scooter providers. 
We use the estimator proposed in \citet{borusyak2021revisiting}, which also straightforwardly allows us to conduct heterogeneity tests for subgroups and to obtain estimates with and without using never-treated cities in our control group.
All our estimates account for period and city fixed effects, and heterogeneous treatment effects under staggered rollout. 

Our data combine administrative traffic accident data from 2016 to mid-2021 at the city-month level with extensive data on the timing of the rollout of e-scooter services by over 30 providers in 93 major European cities. Our estimation sample consists of all cities with a population of at least 100,000 that are not suburbs of larger metropolitan areas, and where a shared e-scooter service was introduced or announced by the end of 2021.
The six countries are selected based on publishing accident data in the required detail and on having launched e-scooter services early enough, to be able to observe medium-run effects.
The launch dates of shared e-scooters are manually collected from public sources and contain data on all firms offering e-scooter services in the six countries (details in \cref{app-data}).

We find that police-reported accidents in cities with shared e-scooter services  increased significantly by \(8.2\pm2.9\%\) (mean estimate~\(\pm\)~std.~error) relative to a counterfactual, estimated from cities where e-scooters launched later.
Treatment effect estimates are larger (\(11.5\pm3.5\%\)) for summer months when e-scooters are used and deployed more intensively, while estimates for winter months are statistically insignificant.
The estimate for the average effect is smaller but still economically and statistically significant (\(4.7\pm2.3\%\)) if the control observations are expanded to include, possibly less comparable, cities that never received e-scooters. Our results are robust to different specifications that address potential concerns related to endogenous treatment timing, or confounding effects of COVID-19 countermeasures and seasonality. 

To explore the underlying mechanisms, we study treatment effect heterogeneity between cities with different traffic and road characteristics.
Following the literature on cycling safety---a closely related mode of transportation---we analyze accidents along the dimensions of separated micromobility-suitable infrastructure (henceforth bike lanes), registered cars per capita, and modal splits. 
In cities with a low density of bike lanes or relatively many registered private cars per capita, the effects are substantial (\(11.5\pm3.9\%\) and \(10.1\pm4.1\%\)). In contrast, we find no significant effect in cities with comparably extensive bicycle infrastructure and smaller effects in cities with a low number of cars per capita. Thus, our heterogeneity results point towards a central role of separated cycling infrastructure and public policy in mitigating accidents. This has important policy implications since many cities, regions, and countries have made political commitments to increasing their micro-mobility modal share (including the use of bicycles, e-scooters, and similar modes of transportation) in the face of climate change.\footnote{E.g., in the Pan-European Master Plan for Cycling, signed by 54 European nations, 16 nations have approved national cycling strategies stating an increased cycling modal share as explicit measurable objectives \citep{ecf}.} The lessons from the changes in modal shares induced by the staggered rollout of shared e-scooter services can serve as a precedent for how to safely increase micro-mobility in cities.

Our results naturally account for substitution between modes of transportation. Raw numbers of e-scooter accidents, even if they were recorded consistently and published for many cities, would be ill-suited to inform policy, because they cannot account for substitution and indirect effects. If, e.g., e-scooters substitute for cars, then cities may see an increase in e-scooter accidents but a reduction in accidents involving cars. The resulting overall effect could be a reduction in accidents, in spite of an increase in e-scooter accidents.
Therefore we study accidents involving personal injury, irrespective of the involved vehicles. The fact that we find large significant treatment effects, suggests that substitution effects must be negligible.

Another reason to focus on overall accidents is that not all accidents caused by e-scooters must necessarily involve an injured e-scooter user. For example, the analysis of two Swedish data sets of injuries associated with e-scooters \citep{stigson2021electric} showed that 8\% of such injuries were sustained by other road users. An additional 5\% were pedestrians or cyclists injured by parked e-scooters.
If shared e-scooters indirectly affect accidents among other modes of transportation, studying overall accidents can reveal those effects regardless of whether detailed classifications of all indirectly involved parties are available.

The existing evidence on injuries caused by e-scooter accidents is of limited generalizability and not suited to inform public policy for two more reasons. First, previous studies are mostly descriptive and based on hospital data in single cities or countries. These analyses usually characterize accident risk factors and types of injuries \citep{badeau2019emergency, sikka2019sharing, trivedi2019injuries, blomberg2019injury, yang2020safety, namiri2020electric, stigson2021electric}, but do not quantify the effects of e-scooter services on overall accidents resulting in injuries.
Second, existing literature does not study the role of different city characteristics in mitigating accidents. We add to the literature on factors affecting and mitigating traffic accidents, such as traffic regulations \citep{peltzman1975effects, abouk2013texting, van2015optimal, bauernschuster2022speed} or road characteristics \citep[for an overview see][]{wang2013effect}.

%% file: MaterialsMethods.tex
\subsection{Data sources and sample}
We limit our study to cities of at least 100,000 inhabitants that are not suburbs of larger cities and the period between January 2016 and June 2021.\footnote{The panel horizon of 6/2021 addresses a trade-off:  Shorter horizons reduce the available post-treatment periods to estimate effects. Longer horizons include periods for which few yet-to-be-treated cities remain to estimate the period fixed effects. Using 6/2021 ensures enough yet-to-be-treated cities remain (see \cref{fig:launch_dates}) without omitting the whole summer of 2021. Estimates are comparable when using different cut-offs, e.g., 12/2020 or 9/2021.} \Cref{app-data} describes all data sources in detail. 

\subsubsection{E-scooter service rollout}
The data on the launch dates of e-scooter services in different cities have been collected from official press releases, social media channels, and websites of e-scooter providers, local newspapers, or cities. We corroborate our data with dates provided directly by two large e-scooter firms. In total, we have data on rollouts for 38 different providers. \Cref{fig:launch_dates} illustrates the recorded launch dates of providers in all included cities until December 2021.

\subsubsection{Traffic accidents}

We have monthly administrative data on traffic accidents involving personal injury for January 2016 to June 2021. Data from Sweden is not available for 2016 and 2017.
Austria, Germany, and Switzerland report traffic accidents by city, however the Scandinavian countries report accidents by municipalities. For large cities, as in our sample, municipalities and cities are well-aligned: the population of the average Scandinavian sample city,  makes up 88\% of corresponding municipality's population (details in \cref{section:traffic_data}).  For the cities of Stockholm, Oslo and Helsinki, which span multiple municipalities, we use data from the homonymous municipality, covering the majority of the respective city's population.

\subsubsection{Heterogeneity variables}
To investigate treatment effect heterogeneity along different city characteristics, we use three different variables: share of separated bike lanes in the road network, registered cars per capita, and cycling modal share. The length of separated cycling infrastructure and total road network length are collected from \emph{OpenStreetMap}. Data on cars per capita and on cycling modal shares by city are primarily obtained from \citet{eurostat}, and supplemented with local administrative sources or academic papers. 

\subsection{Empirical strategy}\label{section:methods}
The analysis of treatment effects under staggered rollout requires special care. Standard multi-period DD estimators---commonly referred to as two-way fixed-effects (TWFE) estimators---can be biased when treatment effects vary over time (e.g., decrease with exposure duration) and when estimation uses data spanning periods in which (i) earlier-treated units remain treated across periods while (ii) later-treated units start being treated.
This bias arises because in standard TWFE models the change in outcomes of earlier-treated groups is used to estimate the counterfactual change over time, in spite of the fact that the outcome of these observations comprises the evolving treatment effect.
Detailed discussions of these issues is found in \citet{goodman2021difference, callaway2021difference, sun2021estimating,borusyak2021revisiting}. 

In our setting the conditions that give rise to the bias of the TWFE estimator fully apply. Treatment timing varies considerably and treatment effects are likely to vary over time for a number of reasons: Fleet sizes grow over time and vary seasonally, urban traffic can adapts to the services, and e-scooter technology improves. For this reason, we base our analyses on two sets of methods that address this issue: i) an event-study estimator based on monthly data \citep{borusyak2021revisiting} and ii) a simple DD estimator based on annual data that uses two groups of cities (early-treated cities and late-treated cities) and compares differences in accidents in 2018 to differences in accidents in 2020. 
All specifications account for city and period fixed effects to capture effects of confounders that are either relatively stable across time (e.g., population, city-specific traffic characteristics, coding or reporting standards) or space (e.g., seasonality in traffic, technological development, climate change). Our results are also robust to allowing for city-specific time trends (\cref{sec:app-r2}).

\subsubsection{Event-study estimates}\label{sec:eventstudy}

For our data, assume outcomes in each city-period are described by  \begin{equation}
    \log \textrm{accidents}_{it} = \alpha_i + \beta_t + \tau_{it}d_{it} + \varepsilon_{it},\label{equation}
\end{equation}
where \(\alpha_i\) is a city-level fixed effect, \(\beta_t\) is a month-level fixed effect,   \(d_{it}\) is a city-month-level indicator for treatment and \(\tau_{it}\) is the city-month-specific treatment effect.
The error term, \(\varepsilon_{it}\), captures random variation in the number of reported accidents that is unrelated to treatment but varies over time and space, e.g., measurement error due to reporting or coding  of accidents.
The individual city-month-specific treatment effect cannot be identified (separately from the error \(\varepsilon_{it}\)).
But we are interested in (conditional) expectations of these effects \(\tau_{it}\), e.g., the average treatment effect across all post-treatment months \(\mathop{\mathbb{E}}[\tau_{it}]\) or the average treatment effect within 12 months of a city's launch date, \(l_i\), \(\mathop{\mathbb{E}}[\tau_{it}|t\leq l_i+12]\). These can be estimated.

Estimation follows three steps. First, only untreated observations (pre-treatment city-months) are used to estimate \(\alpha_i\) and \(\beta_t\). 
Second, these estimated city fixed effects and month fixed effects are used to estimate treatment effects for each treated city-month, \(\widehat{\tau_{it}} = \log \textrm{accidents}_{it} - \hat{\alpha}_i - \hat{\beta}_t\).
Third, these city-month estimates are averaged. 
This average is weighted with weights to match the respective estimand of interest, e.g., the average effect across all post periods (\cref{tab:22did}, col.~1) applies an equal weight to all treated periods and the average effect within 12 months of treatment (\cref{tab:22did}, col.~3) applies an equal weight to the first 12 treatment months by each city and a weight of zero to other observations. The average effects by season (\cref{tab:22did}, col.~4--5) or month since introduction are estimated analogously by applying weights of zero to the respective other months. 

Heterogeneity tests (\cref{sec:mechanisms}) assume accidents are:
\begin{equation}
    \log \textrm{accidents}_{it} = \alpha_i + \beta^\text{h}_t + \beta^\text{h-l}_t\mathbbm{1}_{(x_i>\overline{x}_i)} + \tau_{it}d_{it} + \varepsilon_{it},\label{equationhet}
\end{equation}
where \(\mathbbm{1}_{(x>\overline{x}_i)}\) indicates if a city is above the country median of the specific heterogeneity variable, so \(\beta^\text{h-l}_t\) allows for different month fixed effects depending on the heterogeneity variable. Estimation proceeds as before, estimating city fixed effects and both sets of month fixed effects on control observations only. 
Effect estimates are again obtained as a weighted average of month-city level estimates. 
Tests for equality between effects are conducted by using weights that sum up to 1 for cities above the median and to -1 for cities below the median of the variable of interest, thus estimating \(\mathop{\mathbb{E}}[\tau_{it}|\textrm{above median}] - \mathop{\mathbb{E}}[\tau_{it}|\textrm{below median}]\). This estimate and the corresponding standard error can then be used to test the hypothesis that  \(\mathop{\mathbb{E}}[\tau_{it}|\textrm{above median}] = \mathop{\mathbb{E}}[\tau_{it}|\textrm{below median}]\).

We estimate all models using the natural logarithm of the number of accidents as the dependent variable. 
Effects sizes are likely not constant in absolute terms, but only relative to the baseline number of accidents, e.g., because the number of deployed e-scooters tends to be roughly proportional to city size.
Also, the number of accidents exhibits seasonal swings, with absolute magnitudes roughly proportional to levels. After applying the logarithm, the seasonal swings run parallel across cities, as visualized in \cref{fig:logsparralel}. This effect of the logarithmic transformation is expected if, relative to levels, seasonal swings are parallel across cities. Having parallel seasonality across cities is essential to maintain the parallel trends assumption. 
Using logarithmized dependent variables is unproblematic in our case as observations with zero accidents are rare. Only a single city has zero accidents for only a single month (Oulu, Finland, Feb 2019, two years before e-scooters were introduced). For the estimation, we impute one accident for this observation, but our estimates are qualitatively robust to dropping this observation.

The log-linear specification has the benefit of providing estimates that can be approximately interpreted as semi-elasticities, i.e., the percentage increase in accidents due to the rollout of e-scooter services. 
Tables show transformed estimates expressing non-approximate semi-elasticities \(100\cdot(\exp(\hat{\tau})-1)\%\).

Standard errors allow for the clustering of the model error at the city level.
Standard errors are computed from residuals based on \cref{equation}. To obtain residuals it is necessary to assume a model that is parsimonious enough so that it can separate \(\tau_{it}\) from \(\varepsilon_{it}\). Hence we cannot (as in estimation) allow for arbitrary treatment effect heterogeneity. Instead, we assume that treatment effects are constant within cohorts and use the quarter of scooter introduction to define cohorts, with two exceptions. Zurich, which is the first city that received e-scooters and the only city in Q2 of 2018, is added to the cohort of Q3 2018, and cities that launched scooters in Q1 2021 and Q2 2021 are considered as one cohort. For the heterogeneity analysis in \cref{tab:het_borus}, where the samples are split, several quarters contain too few cities and we thus rely on half-years to define cohorts. To have sufficiently many observations per cohort all cities that launched until June 2019 are considered one cohort and all that launched since December 2021 are considered as one.    
The assumption of constant treatment effects within cohorts bears the risk of being incorrect. However, it was shown that this possible misspecification yields \emph{conservative} inference \citep[][4.3]{borusyak2021revisiting}. Throughout the paper, unless indicated differently, standard errors are computed through a leave-out procedure, that computes the cohort-level treatment effect to obtain the residual, under omission of the focal unit \citep{borusyak2021revisiting}.

\subsubsection{Annual difference-in-differences}
As additional corroboration, we use an annual DD framework comparing changes in accidents from 2018 to 2020 between cities that introduced scooters during 2019 and cities that introduced scooters in or after July 2020. In this panel, there are only two periods and two groups of cities (those that are treated in  2019 and those that are treated later). This model is thus a standard 2-period DD setup, which sidesteps the above-mentioned issues with heterogeneous treatment timing and time-varying treatment effects.

We use 2018, the last year before companies started introducing e-scooter services in most cities as the pre-treatment period, and 2020 as the post-treatment period. Choosing a later post-treatment year would destroy the ability to maintain a comparable set of control cities that are yet-to-be-treated. Choosing an earlier post-treatment year would be disadvantageous because in most sample cities scooters were only introduced during 2019. 
The regression equation is \[\log \textrm{accidents}_{it} = \alpha_i + \delta_t + \beta \textrm{treatment}_{it} + \varepsilon_{it},\]
where \(\textrm{treatment}_{it}\) measures if city \(i\) launched shared e-scooters before period \(t\). \(\alpha_i\) and \(\delta_t\) are city and year fixed effects.

%% file: Results.tex
\Cref{tab:22did}, column 1 reports the average treatment effect across all treated city-months. Accidents involving personal injuries increased on average by \(8.2\pm2.9\%\) (mean estimate~\(\pm\)~std.~error).  To put the estimated effect into perspective: the median sample city in terms of accidents (Potsdam, Germany, population of 180,334) reports 54 accidents in the average pre-treatment month. An increase of 8.2\% thus implies an additional 4.4 monthly accidents. 

\begin{table*}[htbp]
    \caption{Estimated treatment effects on police-reported accidents involving personal injury}\label{tab:22did}
     \adjustbox{max width=\textwidth}{\input{tables/22did.tex}}
     \begin{tablenotes}[flushleft]\footnotesize
    \item \emph{Notes:}
        * p<0.1, ** p<0.05, *** p<0.01. 
        The table shows estimated treatment effects from log-linear specifications (see \cref{section:methods}). Raw estimates are transformed to semi-elasticities: \(100\cdot(\exp(\hat{\tau})-1)\%\). Standard errors are transformed correspondingly.    Col.~1 and 3--7 rely on yet-to-be-treated observations as controls.    Col.~2 and 8 additionally use never-treated cities.     Standard errors are in parentheses. For event-study estimates in col.~1--6, standard errors allow for clustering of the model error at the city level and are computed using the leave-out procedure recommended in \citet{borusyak2021revisiting}, defining cohorts as the quarter in which scooters were launched. For the  estimates in col.~7--8 standard errors are clustered at the city level.   All estimates account for period fixed effects and city fixed effects. 
     \end{tablenotes}
 \end{table*}

Unless stated otherwise, the set of control observations consists of all pre-treatment months for cities where, as of 2021, e-scooters services were announced or introduced. We view these yet-to-be-treated cities as the most comparable counterfactual.
However, the low number of yet-to-be-treated cities towards the end of the sample may be a concern that we address. Column 2 presents estimates using an extended set of control cities that includes never-treated cities where no e-scooter firm launches occurred until 2021. In this sample, the estimated treatment effect is an average increase in total accidents of \(4.7\pm2.3\%\). 
Since this sample is larger, it may allow for a more precise estimation of the counterfactual. However, never-treated cities may be less comparable to (eventually) treated cities. Indeed, never-treated cities in our data are regionally clustered: 15 out of the 17 never-treated cities are German cities, 9 of which are from one state (North Rhine-Westphalia). Including these cities skews the counterfactual towards North Rhine-Westphalia. In addition, most of the never-treated cities are part of a larger metropolitan areas, straddling the line between urban and suburban areas, which are less comparable to the other cities in the sample.
We thus consider the estimates in column~1 more reliable.
An alternative way to address the low number of control cities towards the end of the panel is to rely on yet-to-be-treated cities and shorten the time frame of the analysis by ending earlier.
Re-estimating the specification from column~1 on  a sample ending in 2020 yields an estimate of \(5.8\pm2.3\%\) (\cref{table:sdid}, col.~4).
\Cref{fig:launch_dates} provides an overview of treatment dates, which shows the remaining yet-to-be-treated cities in each period.

The estimates in the first two columns can be interpreted as the average effects  across all post-treatment periods. For the average treated city, the data span 18 post-treatment months. Since all city-months are weighted equally, early-launch cities (spanning up to a maximum of 38 treatment months) have a larger overall weight in the estimate in columns~1--2 than late-launch cities. This is addressed in column~3.
Column 3 reports a short-run effect estimate for the first twelve months after rollout in each city. The average estimated increase in accidents within the first twelve months is smaller than the estimate in column 1, but still statistically and economically significant. 
On the one hand, this difference could be explained by early-adopting cities having larger effects. On the other hand, it could be that effect sizes increase with exposure. 
While a conclusive statement is not possible, \cref{tab:earlyorlong} shows that effect estimates for the second twelve-month period tend to be larger than for the first twelve-months period. First-year effect estimates for early-adopting cities are, however, marginally smaller than estimates for late-adopting cities. This may indicate that effect sizes are increasing in the medium term. But since cities with multiple years of exposure are few, this finding is only suggestive.

Columns 4 and 5 of \cref{tab:22did} show average effects during non-winter months (March--October) and during winter months (November--February), respectively. 
We expect the treatment effect to be concentrated in the non-winter months, since providers tend to reduce the number of deployed e-scooters and e-scooters utilization drop considerably during winter, according to descriptive analyses and news reports  \citep{mathew2019impact, obrien_2021}.  The estimates indicate that e-scooter services significantly increase the total number of accidents by \(11.5\pm3.5\%\) in the non-winter months (col.~4) and only by \(1.9\pm3.3\%\) in the winter months (col.~5), which is statistically indistinguishable from 0.

\begin{figure}[htbp]
     \begin{center}
         \caption{Monthly treatment effects and pre-trends}\label{fig:borusyakplot}
         \includegraphics[width=0.7\linewidth,trim=0.4cm 0 0 1.3cm,clip]{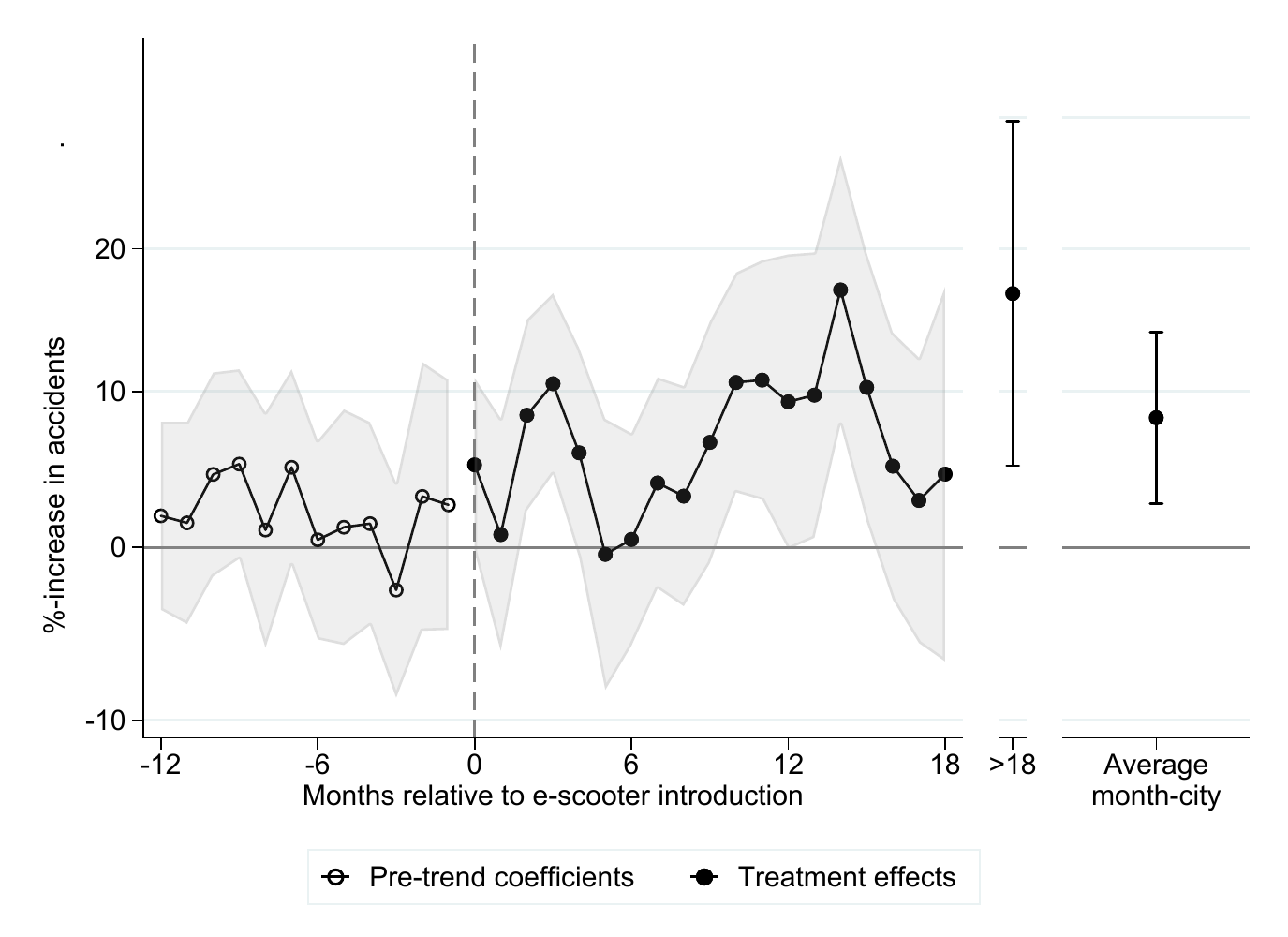}
     \end{center}
     \begin{tablenotes}[flushleft]\footnotesize
        \item \emph{Notes:}
        The figure shows average treatment effects relative to treatment introduction. In the line plot, filled dots indicate treatment effect estimates for the first 18 months relative to the introduction of e-scooter services. Circles indicate estimates for pre-treatment months. Effects after month 18 are combined into a single coefficient, because month-level estimates for long-term effects are estimated on small subsamples  (few cities had e-scooters early enough for long-term effects to materialize). So, monthly estimates for later months cannot be estimated with comparable precision as for earlier months.     In the right part, the average treatment effect estimate and corresponding 95\%-confidence interval, across all available post-treatment city-months is shown (\cref{tab:22did}, col.~1).     Shaded areas/bars indicate the 95\%-confidence interval around the estimates, based on standard errors that allow for the clustering of the model error at the city level, computed via the leave-out procedure using half-year of scooter launch as cohorts (see \cref{section:methods}).     All estimates account for city  fixed effects and month fixed effects.
    \end{tablenotes}
 \end{figure}

A related pattern is observed when we look at average treatment effects by month since the city-specific e-scooter introduction.
\Cref{fig:borusyakplot} shows these estimates, for the first 18 months after the introduction and pre-treatment estimates for the 12 months before.
Post-introduction months consistently indicate increased accident numbers with some apparent heterogeneity in the treatment effect over time.
The fact that the effect of e-scooters on accidents temporarily drops after five months can be explained by a majority of launches occurring in spring and summer. Accordingly, for many cities, the fifth month coincides with the beginning of winter, when we expect less strong effects.
Similarly, the drop around month 17 may be related to the second winter a year later. In line with this conjecture, treatment effect estimates are more stable over time if we exclude winter months from the month-level analysis (\cref{fig:borusyakplotsummer}).

Estimated pre-treatment coefficients are statistically insignificant, suggesting that there was no anticipation effect and no pre-existing differential trends. Additional balance tests and a discussion of (the lack of) pre-treatment differences are in \cref{sec_endog},  \cref{section:app-endogeneity}, and \cref{fig:shiftplacebo}.

To address possible concerns that the treatment effect estimates may be driven by developments related to the COVID-19 pandemic, \cref{tab:22did}, column 6, shows the average effect across all months except the months with the most relevant COVID-19 countermeasures in the sample countries (March to May 2020 and November 2020 to May 2021).
The average treatment effect estimate for the remaining months is \(5.7\pm2.1\%\).

Column 7 and 8 use annually aggregated data and show estimates for the annual DD specification (see \cref{section:methods}). This estimation sample has two groups of cities.
The treatment group consists of all cities where e-scooters were launched in 2019.
For column 7, the control group consists of all yet-to-be-treated cities that were not treated by July 2020.
For column 8, the control group additionally includes never-treated cities, analogous to column 2.
The sample of cities used for in these annual DD estimates is smaller than our main sample but allows for simpler analyses and is thus shown as a corroboration.
The estimates are comparable in magnitude to the  event-study estimates in columns 1-2.

\subsection{Mechanisms}\label{sec:mechanisms}
To investigate mechanisms, we study treatment effect heterogeneity along different traffic and road characteristics of cities. Regulations for e-scooters differ slightly across countries (e.g., Scandinavian countries legally treat the e-scooters in questions as bicycles, while Austria, Germany, and Switzerland, consider e-scooters a separate vehicle category). However, the rules regulating cycle path use are the same for e-scooters as for bicycles \citep{fersi2020}.

Because e-scooters share several characteristics with bicycles, including applicable traffic rules and travel speed, city characteristics associated with cyclist safety are also likely relevant to the effect of shared e-scooters on traffic accidents.
There is also a `safety in numbers'-theory that could extend to e-scooter users, implying that e-scooters could be safer in cities with larger numbers of cyclists \citep{Jacobsen205}.
We thus use three heterogeneity variables related to cyclist safety, which were previously used in related work \citep{kraus2021provisional}: (i) share of bike lanes in the total road network, (ii) the number of registered cars per capita, and (iii) the cycling modal share. The three measures provide complementary evidence as they are collected from different sources and capture constraints faced and choices made by people when choosing their mode of transportation. Infrastructure data is collected from \emph{OpenStreetMaps} \citep{OpenStreetMap}, while data on registered cars comes from national administrative bodies. Modal share estimates are collected from different municipal or independent academic surveys, so they have a higher risk of being measured inconsistently.\Cref{section:het_data} describes the data sources in detail.

For each of the three dimensions, we classify cities into two groups depending on whether they fall above or below the country-specific median. We use country-specific median splits to address concerns about differential reporting standards (e.g., where cars are typically registered) and structural or cultural differences across countries (e.g., what types of cycling infrastructure are typically built). 

 \begin{table}[htbp]
     \caption{Treatment effects are heterogeneous}\label{tab:het_borus}
     \begin{center}
         \input{tables/het_borus.tex}
     \end{center}
          \begin{tablenotes}[flushleft]\footnotesize
     \item \emph{Notes:}
        * p<0.1, ** p<0.05, *** p<0.01. 
        \Cref{tab:het_borus} shows estimated treatment effects from log-linear specifications (see \cref{section:methods}). Standard errors in parentheses. In addition to city fixed effects and month fixed effects, regressions control for interaction fixed effects between month and the median-split indicator. Raw estimates are transformed to semi-elasticities: \(100\cdot(\exp(\hat{\tau})-1)\%\). The table illustrates the average effect estimates for different subsamples, defined by sample splits using the country-level medians of different city characteristics shown in the table header.  Standard errors allow for clustering of the model error at the city level and are computed using the leave-out procedure recommended in \citet{borusyak2021revisiting}, defining cohorts as the half-year in which scooters were launched. Tests for coefficients differences are described in \cref{sec:eventstudy}.
   \end{tablenotes}
 \end{table}
\Cref{tab:het_borus} shows separate treatment effect estimates for the groups defined by each variable's country-specific median split. 
These estimates should not be interpreted as causal effects of the heterogeneity variables themselves, but rather as estimates of the group-specific treatment effect of introducing e-scooters. While the effect of shared e-scooters on accidents in each subsample is identified under the same assumptions as our main effect, the differences in effects between subsamples are not necessarily caused by these heterogeneity variables. Other variables that are correlated with the heterogeneity variables, such as cities' road conditions, average vehicle speeds, speed limits, or income levels, could drive the observed heterogeneity in treatment effects. 

Estimates in column 1 imply that e-scooters only had a small and statistically insignificant effect on total accidents in cities with an above-median density of bike lanes. In contrast, the implied effect for cities with a below-median density of bike lanes is large and highly significant, \(11.5\pm3.9\%\). The difference between the coefficients is highly significant (\(p\)-value=0.018). These results are in line with findings summarized in a recent literature survey \citep{reynolds2009impact}, suggesting that purpose-built bicycle-specific infrastructure can reduce traffic accidents.

The results in column~2 indicate that cities with an above-median density of registered cars experienced a comparatively large treatment effect of \(10.1\pm4.1\%\), while cities with a below-median density of registered cars experienced an estimated increase of \(4.7\pm2.6\%\). 
When splitting the sample of cities by their cycling modal share, the two groups have numerically similar and statistically indistinguishable (\(p\)-value=0.658) estimated effects.  So we find no evidence for the `safety in numbers'-theory for shared e-scooters. The heterogeneity results are similar in the annual DD specification, where we find a similarly large but insignificant difference based on bike lanes and a large and significant difference for cars per capita (\cref{tab:het_did}).

\subsection{Endogeneity, robustness checks, and placebo tests}
\label{sec_endog}
The rollout of scooter services is not random. There are a number of factors that are predictive of market entry. These factors can be relatively time-invariant city characteristics, such as population size, and infrastructure or they can be time-varying, such as season, regulatory constraints, and economic or technological developments. The factors that we identify as drivers of e-scooter launches are either unlikely to cause changes in accidents themselves or are controlled for through the fixed effects in our specifications. We provide a discussion of these factors in this section and in \cref{section:app-endogeneity}.

A threat to the identification of causal effects would be if the city-level timing of launches was correlated with changing trends in accidents, or the reporting and coding of accidents. Accordingly, a key identifying assumption of our empirical strategy is the parallel trend assumption, i.e., total traffic accidents in treatment and control cities would have followed a common trend in the absence of e-scooter services. While this assumption cannot be directly tested, it is more plausible if trends between treated and control cities are similar before the introduction of shared e-scooter services. The estimated pre-trend coefficients are insignificant and close to zero (see \cref{fig:borusyakplot}). As an additional test, we test if past trends in accidents predict the launch of e-scooter services. We regress a city-month level indicator for the launch of e-scooter services on the 12-month difference in log-accidents (\cref{tab:pretrendsonlaunch}) and find no evidence that e-scooter launches are predicted by increases or decreases in accidents. Lastly, placebo tests (discussed below) further substantiate the point that seasonality in accidents or differential trends are not driving our results.

As additional support that endogenous timing of treatment is not driving our findings, we conduct instrumental variable analyses. For this, we use the interactions of four time-invariant city characteristics (population and the three variables used in the heterogeneity analyses above, see \cref{tab:launchpred}) and the number of e-scooter firms that were active in a given month in other cities of the same country as instruments. This analysis, which is carried out in a standard two-way fixed-effects framework---and is thus subject to the discussed caveats related to staggered rollout designs---is discussed and presented in \cref{section:app-endogeneity}. Furthermore, we estimate a synthetic DD model \citep{arkhangelsky2021synthetic}, which slightly relaxes the parallel trends assumption, as an additional robustness check. Our results are qualitatively robust. Details on the synthetic DD are in \cref{app-synth-did}.

More robustness checks and discussions are provided in \cref{app-robustness}. In particular, we provide a detailed discussion of issues arising from the heterogeneous treatment timing in our setting and estimation results for a Poisson model and ordinary least squares (OLS) in a standard (likely biased) TWFE estimation (\cref{tab:twfeiv}). The average effects of shared e-scooter introduction estimated in a standard TWFE DD are qualitatively robust but slightly smaller, as expected.

\subsubsection*{Placebo tests on winter-time accidents and placebo launch dates}
We conduct two types of placebo tests to corroborate our results. First, we use winter-time accidents as a sub-group of accidents that would likely show significant estimates, if our estimates were capturing differential general trends in urban traffic policies or behaviors. E-scooters are a considerably less attractive transportation mode in cold weather and companies tend to reduce the number of deployed scooters \citep{mathew2019impact, obrien_2021}. Therefore, shared e-scooters likely cause fewer accidents in winter. Consequently, the winter specification in column 5 of \cref{tab:22did} can be interpreted as a placebo test. Treatment effects are large and significant in summer months, but smaller and insignificant in winter months. If our results were driven by differential trends in traffic risks related to non-seasonal factors (e.g., car-specific road infrastructure, general policy changes, or changes in the recording and reporting standards) the estimates would be less likely to exhibit this seasonality. 

Second, we redo our estimation from \cref{fig:borusyakplot} with placebo launch dates, shifted by 12 and 24 months to the past from the actual city-specific launch dates. If results were driven by non-parallel trends or by seasonality that is not fully captured by the fixed effects, these placebo tests could also indicate significant estimates. The results are shown in \cref{fig:shiftplacebo} and indicate no discernible patterns, which we view as reaffirming evidence that the parallel trend assumption is justified and that our empirical strategy sufficiently accounts for seasonality.

%% file: tables/22did.tex
{
\def\sym#1{\ifmmode^{#1}\else\(^{#1}\)\fi}
\begin{tabular}{l*{8}{c}}
\toprule
                    &\multicolumn{6}{c}{\shortstack{Monthly event-study estimate\\~}}                               &\multicolumn{2}{c}{\shortstack{2018 vs 2020\\difference-in-differences}}\\\cmidrule(lr){2-7}\cmidrule(lr){8-9}
                    &\multicolumn{1}{c}{(1)}   &\multicolumn{1}{c}{(2)}   &\multicolumn{1}{c}{(3)}   &\multicolumn{1}{c}{(4)}   &\multicolumn{1}{c}{(5)}   &\multicolumn{1}{c}{(6)}   &\multicolumn{1}{c}{(7)}   &\multicolumn{1}{c}{(8)}   \\
                    &               &\shortstack{Incl. never-\\treated cities}   &\shortstack{First 12\\months}   &\shortstack{Non-winter}   &\shortstack{Winter}   &\shortstack{Excluding\\COVID}   &               &\shortstack{Incl. never-\\treated cities}   \\
\midrule
\%-increase in accidents&         8.2***&         4.7** &         5.3** &        11.5***&         1.9   &         5.7***&         9.2***&         6.0** \\
                    &       (2.9)   &       (2.3)   &       (2.1)   &       (3.5)   &       (3.3)   &       (2.1)   &       (2.8)   &       (2.6)   \\
\midrule
Mean pre-treatment accidents&\phantom{space}\clap{93.2}\phantom{space}   &\phantom{space}\clap{83.0}\phantom{space}   &\phantom{space}\clap{93.2}\phantom{space}   &\phantom{space}\clap{100.6}\phantom{space}   &\phantom{space}\clap{78.5}\phantom{space}   &\phantom{space}\clap{95.8}\phantom{space}   &\phantom{space}\clap{1292.1}\phantom{space}   &\phantom{space}\clap{1154.0}\phantom{space}   \\
Treated observations&\phantom{space}\clap{1704}\phantom{space}   &\phantom{space}\clap{1704}\phantom{space}   &\phantom{space}\clap{956}\phantom{space}   &\phantom{space}\clap{1140}\phantom{space}   &\phantom{space}\clap{564}\phantom{space}   &\phantom{space}\clap{948}\phantom{space}   &\phantom{space}\clap{48}\phantom{space}   &\phantom{space}\clap{48}\phantom{space}   \\
Total observations  &\phantom{space}\clap{5880}\phantom{space}   &\phantom{space}\clap{7134}\phantom{space}   &\phantom{space}\clap{5880}\phantom{space}   &\phantom{space}\clap{5880}\phantom{space}   &\phantom{space}\clap{5880}\phantom{space}   &\phantom{space}\clap{5880}\phantom{space}   &\phantom{space}\clap{150}\phantom{space}   &\phantom{space}\clap{188}\phantom{space}   \\
Cities              &\phantom{space}\clap{93}\phantom{space}   &\phantom{space}\clap{112}\phantom{space}   &\phantom{space}\clap{93}\phantom{space}   &\phantom{space}\clap{93}\phantom{space}   &\phantom{space}\clap{93}\phantom{space}   &\phantom{space}\clap{93}\phantom{space}   &\phantom{space}\clap{75}\phantom{space}   &\phantom{space}\clap{94}\phantom{space}   \\
\bottomrule
\end{tabular}
}

%% file: tables/het_borus.tex
{
\def\sym#1{\ifmmode^{#1}\else\(^{#1}\)\fi}
\begin{tabular}{l*{3}{c}}
\toprule
                    &\multicolumn{1}{c}{\shortstack{Share of\\bike lanes}}&\multicolumn{1}{c}{\shortstack{Cars per\\capita}}&\multicolumn{1}{c}{\shortstack{Cycling\\modal share}}\\\cmidrule(lr){2-2}\cmidrule(lr){3-3}\cmidrule(lr){4-4}
                    &\multicolumn{1}{c}{(1)}   &\multicolumn{1}{c}{(2)}   &\multicolumn{1}{c}{(3)}   \\
\midrule
\%-increase for above-median cities&         0.2   &        10.1***&         9.6** \\
                    &       (2.7)   &       (4.1)   &       (4.7)   \\
\addlinespace
\%-increase for below-median cities&        11.5***&         4.7*  &         7.3***\\
                    &       (3.9)   &       (2.6)   &       (2.6)   \\
\midrule
\(p\)-value \(H_o\): coefficients identical&       0.018   &       0.256   &       0.658   \\
Cities              &          93   &          93   &          93   \\
Observations        &        5880   &        5880   &        5880   \\
\bottomrule
\end{tabular}
}

%% file: Discussion.tex
We provide evidence that the rollout of shared e-scooter services in large cities led to a significant increase in traffic accidents involving personal injuries. In their 2018 road safety report, the \citet{oecd2018} reports that the estimated socio-economic costs of road traffic accidents exceed €500bn for EU member states alone, which is equivalent to around 3\% of the EU's GDP. In 2019, an average traffic accident involving personal injuries in Germany was associated with economic costs of around €61,000, including costs of €16,301 for damage to property and €44,778 for personal injuries \citep{bast2021accident,bast2021cost}.\footnote{According to the German federal research institute,  \citet{bast2021cost}, 300,143 traffic accidents involving personal injuries were recorded in Germany in 2019. The total personal damage costs amounted to €13.44bn, which implies average personal damage costs of around €44,778 per accident involving personal injuries.} Assuming these costs per accident apply to  all six sample countries, the estimated effect in our main specification (\(8.2\pm2.9\%\)) would thus imply additional socio-economic costs of around €466,186 per month and €5.6m per year for the average sample city with 93.2 monthly accidents before treatment (see \cref{tab:22did}).\footnote{The true costs may be higher, as Germany does not account for under-reporting in cost calculations \citep{wijnen2017crash}, or lower as the cost estimates also include non-urban traffic accidents.}

In addition, a recent study conducted at the University Hospital of Essen in Germany \citep{meyer2022scooter} suggests that a large share of hospital-treated e-scooter injuries are not reported to the police. By focusing on police-reported accidents our data is more likely to record accidents involving automobiles than accidents involving cyclists and includes relatively more accidents with severe injuries and larger costs \citep{Langley376}. Findings from retrospective studies examining medical records further show that e-scooter-related accidents are often associated with serious injuries to the head and upper extremities with a substantial proportion of major trauma injuries \citep{trivedi2019craniofacial, trivedi2019injuries, badeau2019emergency, moftakhar2021incidence, lavoie2021characterization}. In sum, these findings imply that our estimated effects are associated with significant socio-economic costs that have to be weighed against potential benefits from adding shared e-scooter services to the urban transportation landscape.

Our main results do not differentiate by severity of injuries.
Four countries report the number of accidents involving severe or slight personal injuries and death. Austria and Norway instead disaggregate the number of victims by injury severity.
We do not suspect that e-scooter-related accidents are considerably less severe than other urban accidents for four reasons.
First, the clinical studies discussed above suggest that police-reported accidents directly involving e-scooters often involve severe injuries.
Second, police-reported accidents indirectly caused by e-scooters are likely similar to pre-existing urban accidents.
Third, in the appendix we report estimates of treatment effects on accident severity. For this we use the percentage of accidents involving only slight injury as an outcome variable.
For Austria and Norway we infer the percentage of slight-injury accidents from the percentage of slight-injury victims. 
We find that 85\% of accidents involving personal injuries are classified as involving slight injuries and that there is no significant change after the introduction of e-scooters (95\% confidence interval:  \(-1.8\) to \(+0.8\)\%-points, details in \cref{sec:app-r1}).

Estimated effects are considerably larger in cities that have poor separated cycling infrastructure and rely more on cars. This is important from an urban planning policy perspective, 
especially given that earlier studies suggest that better bicycle infrastructure may be associated with more frequent and longer e-scooter trips \citep{caspi2020spatial, laa2020survey}. Paired with our findings this suggest that improving separated infrastructure may curb negative effects, while simultaneously encouraging e-scooter usage. 
As highlighted in the results section, the effects of infrastructure are not causally identified. Investigating what additional measures those cities that have well-developed infrastructure are taking to prevent accidents is worth additional study.

Our analyses focus on the extensive margin of treatment, i.e., whether or not shared e-scooters are available in a city at all. This is the most relevant dimension for two reasons. First, while there are examples of cities that restricted the maximum number of e-scooters (e.g. Bern, Switzerland; Innsbruck, Austria), the public debate usually revolves around the extensive margin, i.e., whether to ban shared e-scooters from a city's transportation mix or not \citep{cnn_ban}. 
Second, the extensive margin analysis allows for an easy interpretation as a semi-elasticity and is less likely to be subject to measurement error and endogeneity issues. Even if exact data on the number of deployed e-scooters were available, e-scooter firms may continuously adapt to accidents and changing traffic conditions, raising concerns about reverse causality. Extensive margin estimates, on the other hand, identify a net effect that includes possible endogenous scaling decisions of suppliers.

Our findings should not be interpreted as long-run effects. Effects on traffic accidents may decrease in the future due to safer vehicles, greater experience of users of e-scooters, or changes in the behavior of other traffic participants. They may also increase. While our analyses account for substitution effects between modes of transportation, including walking (police are required to report pedestrian-vehicle accidents), we do not observe the total number of trips within a city. If the introduction of shared e-scooters considerably increases the mobility of citizens in treated cities, i.e., people travel more because e-scooters are available, the social costs, implied by documented effects, should be discounted by the higher number of trips in a cost-benefit analysis. A recent survey of e-scooter users in Paris, however, suggests that the vast majority did not increase their total mobility \citep{christoforou2021using}. 

Our analysis does not imply a comparative statement of the safety risks between different modes of transport and is agnostic to what kind of road user is responsible for the increased accidents. Consequently, our results cannot be interpreted as recommendations against the inclusion of shared e-scooters in urban transport landscapes, in particular, compared to automobiles that, due to their size and speed, likely pose the relatively largest urban safety risk. For instance, past studies find that cars and other large motorized vehicles contribute to other road users' deaths at rates 3--6 times higher than bicycles per mile driven \citep{aldred2021does, scholes2018fatality}.

%% file: app-robustness.tex
\subsubsection{Why the two-way fixed effects estimator is likely biased}\label{sec:app-poisson}
Our main specifications account for biases that may arise due to heterogeneous treatment timing and heterogeneous treatment effects that were studied extensively in the recent econometric literature on two-way fixed effects (TWFE) difference-in-differences (DD) estimators \citep{borusyak2021revisiting,goodman2021difference,sun2021estimating,callaway2021difference}. 

The standard TWFE DD estimate for the treatment effect (also as a Poisson regression) is likely downward biased in our setting. To understand this, note that standard DD estimation is based on comparing the average change in units that switch their status from untreated to treated in a given period to the average change in units with no change in treatment status. For multi-period staggered-rollout designs, the group with no change in treatment status consists of two subgroups: untreated units and units that were treated in earlier periods and remain treated. If the treatment effect on the earlier treated units changes over time the earlier treated units do not provide a valid counterfactual because the evolution of their outcome combines the counterfactual change/trend and the changing treatment effect.

In our setting, a changing treatment effect is expected, e.g., because e-scooter companies gradually increase the number of deployed scooters, because commuters only gradually start using them, and because traffic gradually adapts to the presence of e-scooters. The evolution of treatment effects over time is neither the focus of this research nor robustly identifiable from currently available data. But given that the number of deployed e-scooters typically increases over time and that adoption is likely gradual, it seems plausible that treatment effects increase over time. This implies that estimation should not be based on the assumption of constant treatment effects as this would bias TWFE DD estimates downward.

For comparison, OLS estimates of standard TWFE DD regression and a Poisson regression estimate of the same model (that we will discuss in \cref{sec:app-poisson2}), which do not account for this heterogeneity can be found in columns 3, 4, 11, and 12 of \cref{tab:twfeiv}.

\begin{table*}[hbpt]
    \caption{Country-time fixed effect regressions and additional specifications to study extensive vs intensive margin effects and the possibility of endogenous placement of scooter firms.}
    \label{tab:twfeiv}
    \centering
    \adjustbox{max width=\textwidth}{
        \input{tables/twfe2sls.tex}
    }
        \begin{tablenotes}[flushleft]\footnotesize
    \item \emph{Notes:}  * p<0.1, ** p<0.05, *** p<0.01.
    Robust standard errors in parentheses allow for clustering of the model error at the city level.
    Raw estimates are transformed to semi-elasticities: \(100\cdot(\exp(\hat{\tau})-1)\%\).
    Estimates for treatment effects are based on different specifications and estimators as indicated in the table header. Even columns additionally account for year-country fixed effects.
    Column 1 show estimates from our main specification (\cref{tab:22did}, column1) for convenience.
    The estimate in column 2 accounts for country-year fixed effects and excludes observations for Austria 2020-21 and Norway 2021, because for those countries there exist no ``later-treated'' cities to identify the country-year fixed effects from. All cities that meet the sample criteria were treated.
    Columns 3--4 show TWFE estimates.
    Columns 5--6 show instrumental variable estimates as discussed in \cref{section:app-endogeneity}. The Kleibergen-Paap Wald F-statistics tests for the exclusion of the instruments in the first stage of the instrumental variable.
    Columns 7--10 follow the same order as columns 3--6, but use the count of scooter firms, as opposed to the launch indicator for the first scooter firm, as the main independent variable. Columns 11--12 show estimates from a TWFE Poisson regression.
\end{tablenotes}
\end{table*}

Including more granular fixed effects (i.e., country-year) in a TWFE DD framework may partially reduce concerns about these biases (e.g., country-year fixed effects would avoid comparing later treated German cities against earlier treated Swedish or Austrian cities). Indeed, \cref{tab:twfeiv} illustrates that including such fixed effects renders estimates from TWFE DD, based on the log-linearized model or a Poisson model, consistently either larger or more precisely estimated.

However, using country-year fixed effects is not a satisfactory remedy, as the described problem still exists for the implied comparisons within countries. More importantly, the inclusion of country-year fixed effects removes all variation stemming from countries and years where all cities are treated (i.e., Austria 2020--2021 and Norway in 2021). The estimand is thus only the average treatment effect in the remaining years and countries---especially if treatment effects grow dynamically, this restriction implies losing relevant years of possibly significant effects. The imputation estimator that we use addresses the issue of constructing a counterfactual that is not affected by dynamic treatment effects at its root by estimating the fixed effects from control observations only, which avoids any implicit comparisons of newly treated observations to earlier treated observations.

\subsubsection{Poisson versus logarithmized dependent variables}\label{sec:app-poisson2}

We estimate the effect of introducing shared e-scooter services in our main model as a semi-elasticity by applying the natural logarithm to the dependent variable in a linear model that we estimate with OLS, which can be problematic when the dependent variable contains many zeros.
In principle, semi-elasticities could also be estimated through a Poisson regression without transforming the dependent variable.
We rely on the log transformation for two reasons. First, observations with zero accidents are extremely rare in our data (only a single month-city in our sample has recorded zero accidents, as pointed out in \cref{section:methods}). Second, to our knowledge, methods for Poisson regression that account for treatment effect heterogeneity in staggered rollout settings have not been developed. 

To investigate whether using a log transformation of the dependent variable, as opposed to modeling a Poisson distribution, significantly changes conclusions in our setting, we estimate a Poisson variant of the two-way fixed effect regression \[\log\mathbb{E}[\textrm{accidents}_{it}] = \alpha_i + \beta_t + \tau d_{it},\]
where \(d_{it}\) is an indicator for treatment in city \(i\) and month \(t\).
As can be seen in \cref{tab:twfeiv}, the estimates from the log-linear DD specifications (e.g., col.~4) and the estimates obtained from the Poisson model (e.g., col.~12) support qualitatively comparable conclusions, especially when country-year fixed effects are included. This similarity reassures us that our main results are not a consequence of relying on a log-transformed dependent variable, as opposed to a Poisson model.

%% file: tables/twfe2sls.tex
{
\def\sym#1{\ifmmode^{#1}\else\(^{#1}\)\fi}
\begin{tabular}{l*{12}{c}}
\toprule
                    &\multicolumn{2}{c}{\shortstack{Event-study estimate\\(Borusyak et al. 2021)}}&\multicolumn{4}{c}{\shortstack{Two-way fixed effects\\(TWFE)}} &\multicolumn{4}{c}{TWFE intensive margin}                      &\multicolumn{2}{c}{Poisson regression}\\\cmidrule(lr){2-3}\cmidrule(lr){4-7}\cmidrule(lr){8-11}\cmidrule(lr){12-13}
                    &\multicolumn{1}{c}{(1)}&\multicolumn{1}{c}{(2)}&\multicolumn{1}{c}{(3)}&\multicolumn{1}{c}{(4)}&\multicolumn{1}{c}{(5)}&\multicolumn{1}{c}{(6)}&\multicolumn{1}{c}{(7)}&\multicolumn{1}{c}{(8)}&\multicolumn{1}{c}{(9)}&\multicolumn{1}{c}{(10)}&\multicolumn{1}{c}{(11)}&\multicolumn{1}{c}{(12)}\\
                    &\multicolumn{1}{c}{}&\multicolumn{1}{c}{}&\multicolumn{1}{c}{OLS}&\multicolumn{1}{c}{OLS}&\multicolumn{1}{c}{IV}&\multicolumn{1}{c}{IV}&\multicolumn{1}{c}{OLS}&\multicolumn{1}{c}{OLS}&\multicolumn{1}{c}{IV}&\multicolumn{1}{c}{IV}&\multicolumn{1}{c}{PPML}&\multicolumn{1}{c}{PPML}\\
\midrule
Effect of introduction (\%)&         8.2***&         4.6** &         4.7** &         4.6***&        14.6** &        15.3***&               &               &               &               &         2.7*  &         4.1***\\
                    &       (2.9)   &       (2.4)   &       (2.0)   &       (1.7)   &       (7.4)   &       (5.4)   &               &               &               &               &       (1.6)   &       (1.3)   \\
\addlinespace
Effect of one additional company (\%)&               &               &               &               &               &               &         1.0** &         1.0** &         2.2*  &         2.2** &               &               \\
                    &               &               &               &               &               &               &       (0.5)   &       (0.4)   &       (1.2)   &       (1.0)   &               &               \\
\addlinespace
Country-year FEs    &           ~   &\(\checkmark\)   &           ~   &\(\checkmark\)   &           ~   &\(\checkmark\)   &           ~   &\(\checkmark\)   &           ~   &\(\checkmark\)   &           ~   &\(\checkmark\)   \\
\midrule
Kleibergen-Paap Wald rk \rlap{\(F\)-stat}&               &               &               &               &         9.1   &        10.1   &               &               &         6.8   &        14.0   &               &               \\
\midrule 
Cities              &          93   &          93   &          93   &          93   &          93   &          93   &          93   &          93   &          93   &          93   &          93   &          93   \\
Observations        &        5880   &        5784   &        5880   &        5880   &        5880   &        5880   &        5880   &        5880   &        5880   &        5880   &        5880   &        5880   \\
\bottomrule
\end{tabular}
}

%% file: app-endogenity.tex
\subsubsection{Factors that predict treatment timing and why they are unlikely to confound our results}\label{sec:app-predict}

The rollout of scooter services is not random. There are a number
of factors that are predictive of city-level market entry. These factors can be relatively time-invariant, such as population size or infrastructure, or they can be time-varying, such as season and regulatory constraints. In this section, we identify such factors and argue why they are either sufficiently accounted for in our analyses, through the included fixed effects, or how they can be assumed to be unrelated to accident numbers.

Two factors that can be considered time-invariant over our period of study clearly predict the timing of rollouts.
First, firms initially target larger cities. As shown in \cref{fig:launch_dates}, each country's largest cities were among the first to be treated. 
Second, firms prioritize cities that are more bicycle-friendly, where e-scooters are arguably more likely to be successfully adopted. \Cref{tab:launchpred} shows that larger cities and cities with an extensive separated cycling road network received e-scooters significantly earlier.
Cities with 100,000 more inhabitants received e-scooters on average 0.3-0.6 months earlier. Cities with a one standard deviation larger share of bike lanes received e-scooters on average 3-4 months earlier. Similarly, cities with a one standard deviation larger number of cars per capita received e-scooters 3-4 months later.
This is evidence that the timing of launches is to a significant extent driven by time-invariant characteristics of cities.
In our estimation of the treatment effect of e-scooters, all time-invariant characteristics of cities are controlled for through city-level fixed effects.

\begin{table}[htbp]
    \centering
    \caption{Predictors of early launches}
        \begin{center}    
    \adjustbox{max width=\linewidth}{
        \input{tables/launchpred}
    }
    \end{center}    
    \label{tab:launchpred}
        \begin{tablenotes}[flushleft]\footnotesize
    \item \emph{Notes}: The table shows the coefficients of an OLS regression, regressing the month of introduction on different city characteristics. All independent variables, except for population are normalized to have mean 0 and variance 1. Positive [negative] coefficients imply later [earlier] introductions. All regressions control for country fixed effects. Robust standard errors in parentheses.
    \end{tablenotes}
\end{table}

Three time-varying factors that predict rollout timing can also be identified.
First, within a year the start of operations is often in summer when the service is attractive to customers. Roughly half of the launch dates are either in June, July, or August (see \cref{fig:launch_dates}).
Second, firms target cities that are close to recently added areas of operation, likely to exploit economies of scale. For instance, providers usually roll out their services country-by-country. When they decided to expand to a country, they usually quickly roll out their services in all key cities within the respective country \citep[see for example the press statement by Voi on its expansion in Germany,][]{voi2019}. Also within countries, this spatial correlation can be empirically observed in the data.
Between January 2018 and June 2021, in the average sample city, the probability that a new firm launched in a given month was 6.6\%. In the first two [one] months within a new firm launching in the nearest neighboring city (conditional on new launches anywhere else in the country) this probability is higher by around \(5.1\pm1.5\) [\(7.4\pm2.4\)] percentage points than in other months (see \cref{tab:nbsonlaunch}, columns 1 and 2). This pattern is likely driven by firm-level expansion waves, but it also translates into a spatial correlation of the overall e-scooter rollout (see \cref{tab:nbsonlaunch}, columns 3 and 4).
Third, e-scooter launches are subject to regulation, which poses a binding constraint to the timing and is unlikely to independently affect changes in accidents. For example, in Germany, the timing of launches for almost half of the cities coincided with federal regulation in June 2019 that initially allowed the use of e-scooters on public roads \citep{gebhardt2021ll}. Scooter providers were ready to start operations right after the regulation allowed them to.

\begin{table}[htbp]
    \caption{Launches in neighboring cities predict launches of e-scooter firms.}
    \begin{center}    
    \adjustbox{max width=\linewidth}{
        \input{tables/neigboursonlaunch}
    }
    \end{center}    
    \label{tab:nbsonlaunch}
        \begin{tablenotes}[flushleft]\footnotesize
    \item \emph{Notes}: The table shows coefficients (scaled to be interpretable as percentage points) of a regression of indicators for firm launch between 2018 and June 2021 on firm-launch indicators for neighboring cities. All regressions control for city and month fixed effects. Clustered standard errors in parentheses allow for city-level clustering. `Neighboring city' is the geographically closest sample city where e-scooters were ever launched.
    \end{tablenotes}
\end{table}

These three time-varying factors are unlikely to co-determine accidents (e.g., the number of e-scooter firms in a neighboring city can be assumed to not affect accident numbers; scooter-specific regulation can be assumed to not affect accidents through other channels---as individually-owned e-scooters are rare) or are controlled for through the use of month fixed effects (e.g., summer months generally show higher accident numbers).

\begin{table}[htbp]
    \caption{City trends in accidents do not predict launches of e-scooter firms.}
    \begin{center}    
    \adjustbox{max width=\linewidth}{
        \input{tables/pretrendsonlaunch}
    }
    \end{center}    
    \label{tab:pretrendsonlaunch}
        \begin{tablenotes}[flushleft]\footnotesize
    \item \emph{Notes}: The table shows coefficients (scaled to be interpretable as percentage points) of a regression of indicators for firm launch between 2018 and June 2021 on preceding changes in accidents. All regressions control for city and month fixed effects. Clustered standard errors in parentheses allow for city-level clustering.
    \end{tablenotes}
\end{table}

In sum, there are a number of factors determining e-scooter rollout. But the city-specific timing of scooter launches is unlikely to be endogenous to city-idiosyncratic trends in the number of accidents in our empirical model.

\subsubsection{An instrumental variable approach}

As additional corroboration, we use an instrumental variable (IV) approach to address possible endogeneity concerns. Instruments need to vary over time and space. To construct instruments, we use cities' population and our heterogeneity variables, which vary cross-sectionally, and the number of active e-scooter firms in other cities of the same country, which varies over time. The interactions of the former with the latter are used to instrument for treatment in a two-way fixed effects model, accounting for time and unit fixed effects. In particular, for the cross-sectional variation, we use the three pre-treatment variables from the heterogeneity analysis in \cref{tab:het_borus} (demeaned at the country level). For the temporal variation, we instrument the
binary treatment indicator by using a dummy variable for whether in other cities in the same country e-scooters were launched (\cref{tab:twfeiv}, col. 5 and 6). In addition, we also run an intensive margin TWFE-IV specification (\cref{tab:twfeiv}, col. 9 and 10) where we instrument the count of e-scooter providers in a city-month by using the total number of firms that have launched services in other cities in the same country as temporal variation.

For the interaction terms to be valid instruments, they need to fulfill two conditions. First, they need to predict e-scooter rollout, which they do (see \cref{sec:app-predict}). However, weak identification concerns may be valid and the results should be considered estimated with significant uncertainty. The first-stage \(F\)-statistics are reported in the table footer of \cref{tab:twfeiv} column 5, 6, 9, and 10.

Second, the interaction terms between scooter firms and city-characteristics should affect accidents only through the supply of shared e-scooter services. Since the global development of the scooter market and city characteristics are already controlled for through the use of city and month fixed effects, this assumption may be justified. 

The IV results, shown in \cref{tab:twfeiv} columns~5, 6, 9, and 10, indicate large positive and statistically significant treatment effects that are, however, estimated with relatively large standard errors. While the TWFE-IV results reaffirm our main findings, we consider these results only of secondary importance because of possible weak instrument concerns and the inability of the TWFE-IV framework to allow for heterogeneous treatment effects. Most importantly, we do not consider endogeneity concerns a strong concern in our main specification. 

%% file: tables/launchpred.tex
{
\def\sym#1{\ifmmode^{#1}\else\(^{#1}\)\fi}
\begin{tabular}{l*{4}{c}}
\toprule
                    &\multicolumn{1}{c}{(1)}&\multicolumn{1}{c}{(2)}&\multicolumn{1}{c}{(3)}&\multicolumn{1}{c}{(4)}\\
                    &\multicolumn{1}{c}{\shortstack{Month of\\introduction}}&\multicolumn{1}{c}{\shortstack{Month of\\introduction}}&\multicolumn{1}{c}{\shortstack{Month of\\introduction}}&\multicolumn{1}{c}{\shortstack{Month of\\introduction}}\\
\midrule
Population (in 100k)&      -0.608***&      -0.278   &      -0.637***&      -0.301*  \\
                    &     (0.169)   &     (0.192)   &     (0.231)   &     (0.161)   \\
\addlinespace
Share of bike lanes &      -3.925***&               &               &      -3.663***\\
                    &     (0.979)   &               &               &     (1.114)   \\
\addlinespace
Cars per capita     &               &       3.545** &               &       2.951** \\
                    &               &     (1.393)   &               &     (1.417)   \\
\addlinespace
Cycling modal share &               &               &      -0.699   &       0.982   \\
                    &               &               &     (0.831)   &     (0.878)   \\
\midrule
Cities              &          93   &          93   &          93   &          93   \\
\bottomrule
\end{tabular}
}

%% file: tables/neigboursonlaunch.tex
{
\def\sym#1{\ifmmode^{#1}\else\(^{#1}\)\fi}
\begin{tabular}{l*{4}{c}}
\toprule
Dependent variable: &\multicolumn{2}{c}{=1 if new firm launches}&\multicolumn{2}{c}{=1 if first firm launches}\\\cmidrule(lr){2-3}\cmidrule(lr){4-5}
                    &\multicolumn{1}{c}{(1)}   &\multicolumn{1}{c}{(2)}   &\multicolumn{1}{c}{(3)}   &\multicolumn{1}{c}{(4)}   \\
\midrule
Launches in neighbor city (2 months)&         5.1***&               &         1.5*  &               \\
                    &       (1.5)   &               &       (0.8)   &               \\
\addlinespace
Launches in neighbor city (same month)&               &         7.4***&               &         2.0   \\
                    &               &       (2.4)   &               &       (1.4)   \\
\addlinespace
Launches in country (2 months)&         0.7***&               &         0.4***&               \\
                    &       (0.1)   &               &       (0.1)   &               \\
\addlinespace
Launches in country (same month)&               &         1.4***&               &         0.7***\\
                    &               &       (0.3)   &               &       (0.1)   \\
\midrule
Mean dep. var.      &       6.6\%   &       6.6\%   &       2.3\%   &       2.3\%   \\
Cities              &          93   &          93   &          93   &          93   \\
Observations        &        3826   &        3839   &        3826   &        3839   \\
\bottomrule
\end{tabular}
}

%% file: tables/pretrendsonlaunch.tex
{
\def\sym#1{\ifmmode^{#1}\else\(^{#1}\)\fi}
\begin{tabular}{l*{4}{c}}
\toprule
Dependent variable: &\multicolumn{2}{c}{=1 if new firm launches}&\multicolumn{2}{c}{=1 if first firm launches}\\\cmidrule(lr){2-3}\cmidrule(lr){4-5}
                    &\multicolumn{1}{c}{(1)}   &\multicolumn{1}{c}{(2)}   &\multicolumn{1}{c}{(3)}   &\multicolumn{1}{c}{(4)}   \\
\midrule
\(\Delta_{12}\log \textrm{accidents}_{it}\)&        -0.7   &               &         0.1   &               \\
                    &       (1.2)   &               &       (0.9)   &               \\
\addlinespace
\(\Delta_{12}\log(\sum_{k=1}^{12}\textrm{accidents}_{i\{t-k\}})\)&               &        -3.3   &               &        -1.9   \\
                    &               &       (5.6)   &               &       (3.1)   \\
\midrule
Mean dep. var.      &       6.6\%   &       6.4\%   &       2.3\%   &       2.2\%   \\
Cities              &          93   &          92   &          93   &          92   \\
Observations        &        3768   &        3648   &        3768   &        3648   \\
\bottomrule
\end{tabular}
}

%% file: app-more-tables.tex
This section contains a number of tables and figures to support arguments in the main text.

\Cref{tab:earlyorlong} computes various aggregations of the monthly treatment effects to investigate differences between short-term and long-term effects and between effects on early adopters and on late adopters. This is suggestive evidence that effect sizes might be growing over time.

\begin{table}[htbp]
    \caption{Contrasting effect estimates for cities with early or late launches, and for the first or second year}
    \label{tab:earlyorlong}
    \begin{center}    
    \adjustbox{max width=\linewidth}{
        \input{tables/early_or_long.tex}
    }
    \end{center}    
        \begin{tablenotes}[flushleft]\footnotesize
    \item \emph{Notes:}  * p<0.1, ** p<0.05, *** p<0.01.
    The table illustrates the average treatment effect estimates for the first 12 post-treatment months (col 1.), and for the second 12-month period (months 12-13) after the introduction of e-scooters (col. 2). Columns 3 and 4 show the average treatment effect estimates for ``Early'' and ``Late adopters'', considering the first 12 post-treatment months. ``Early adopters'' are cities in which e-scooters were launched on or before August 2018. ``Late adopters'' are cities that launched e-scooters after that.  Standard errors in parentheses are computed using the leave-out procedure recommended in \citet{borusyak2021revisiting} and defining cohorts based on the quarter  in which scooters were launched. All estimations allow for month fixed effects and city fixed effects.
    \end{tablenotes}
\end{table}

\Cref{tab:het_did} repeats the heterogeneity analysis from \cref{tab:het_borus} in the annual DD framework. Qualitatively the results are similar to the results from the monthly framework, estimating the largest effects for cities with a low share of bike lanes and for cities with the largest number of cars per capita.

\begin{table}[htbp]
    \caption{Heterogeneity of treatment effects, estimated in the annual difference-in-differences framework}
    \label{tab:het_did}
    \begin{center}    
    \adjustbox{max width=\linewidth}{
        \input{tables/het_did.tex}
    }
    \end{center}    
        \begin{tablenotes}[flushleft]\footnotesize
    \item \emph{Notes:}  * p<0.1, ** p<0.05, *** p<0.01.
     The table illustrates the average treatment effect estimates according to the annual DD framework for different subsamples that were split subject to the median of different city characteristics. Median splits are based on the country-level medians of each variable among sample cities. Standard errors in parentheses account for clustering at the city level.
    \end{tablenotes}
\end{table}

\begin{figure}[htb]
    \centering
    \caption{Monthly treatment effects and pre-trends for non-winter months}
    \begin{center}    
    \includegraphics[width=0.8\linewidth,trim=0.8cm 0 0 0.5cm,clip]{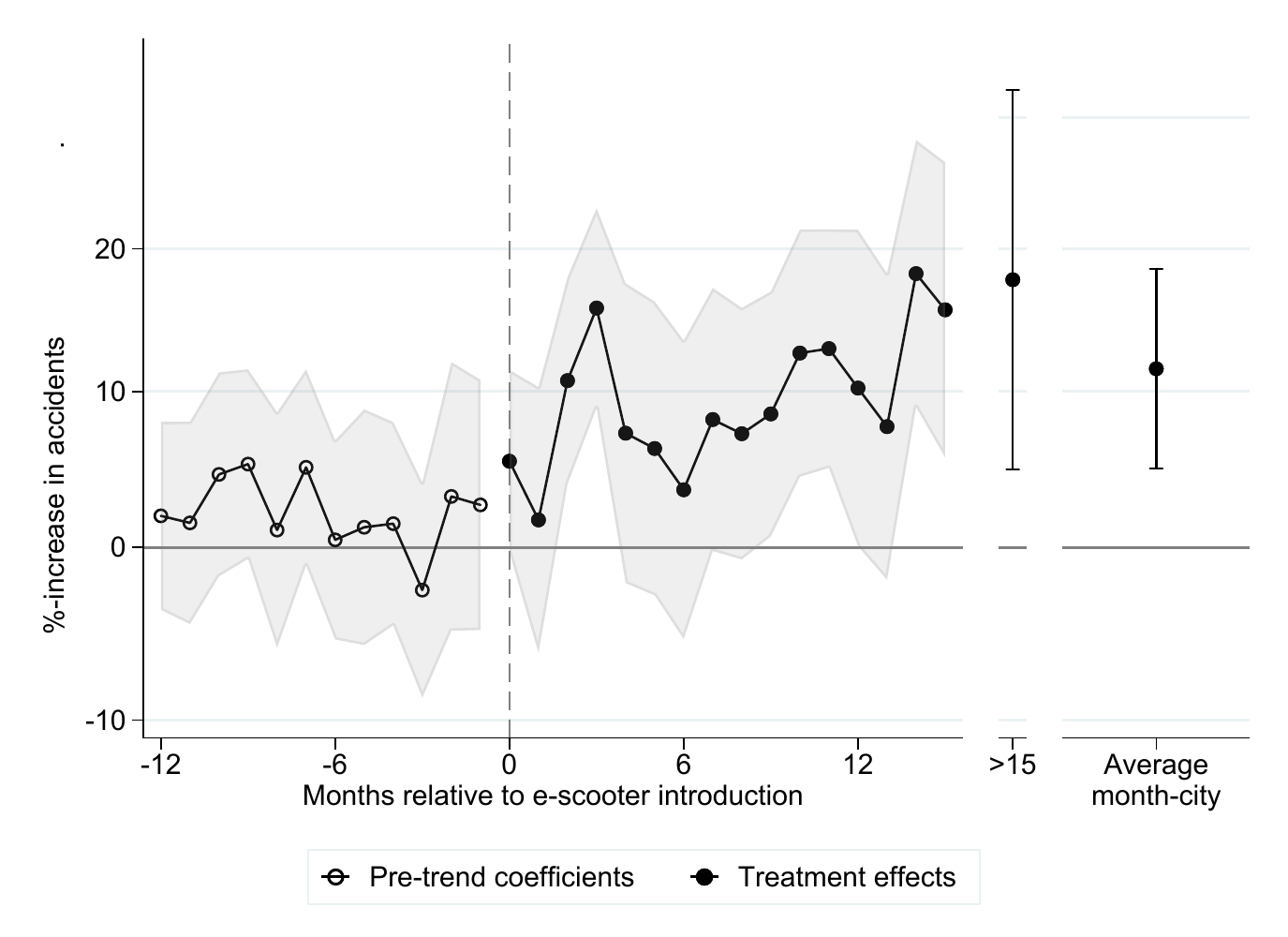}
    \end{center}    
        \begin{tablenotes}[flushleft]\footnotesize
    \item \emph{Notes:} 
    The figure shows average treatment effects relative to the treatment timing. City-months that fall into Nov.--Feb. receive a weight of zero.
    Raw estimates are transformed to semi-elasticities: \(100\cdot(\exp(\hat{\tau})-1)\%\).
    In the line plot, filled dots indicate treatment effect estimates for the first 18 months relative to the introduction of e-scooter services. Circles indicate estimates for pre-treatment months. Effects after month 15 are combined into a single coefficient, because month-level estimates for long-term effects are estimated on small subsamples  (only few cities had e-scooters early enough for long-term effects to materialize). Monthly estimates for later months could thus not be estimated with comparable precision as for earlier months.
    In the right part, the average treatment effect estimate and corresponding 95\%-confidence interval, across all available post-treatment city-months is shown (\cref{tab:22did}, col.~3). 
    Shaded areas/bars indicate the 95\%-confidence interval around the estimates, based on leave-out standard errors defining half-year of scooter launch as cohorts, as described in materials and methods, sub\cref{section:methods}.
    All estimates account for city and month fixed effects.
    \end{tablenotes}
    \label{fig:borusyakplotsummer}
\end{figure}

\Cref{fig:borusyakplotsummer} shows the same analysis as \cref{fig:borusyakplot} but restricts the estimand to summer-month treatment effects. As a result, the apparent dips in treatment effects around observed in \cref{fig:borusyakplot} cease to appear. We take this as evidence for our conjecture that these dips are driven by winter months.

\begin{figure}[htbp]
    \centering
    \caption{Placebo tests using treatment dates shifted by 12 or 24 months.}
    \begin{center}    
    \includegraphics[width=0.8\linewidth]{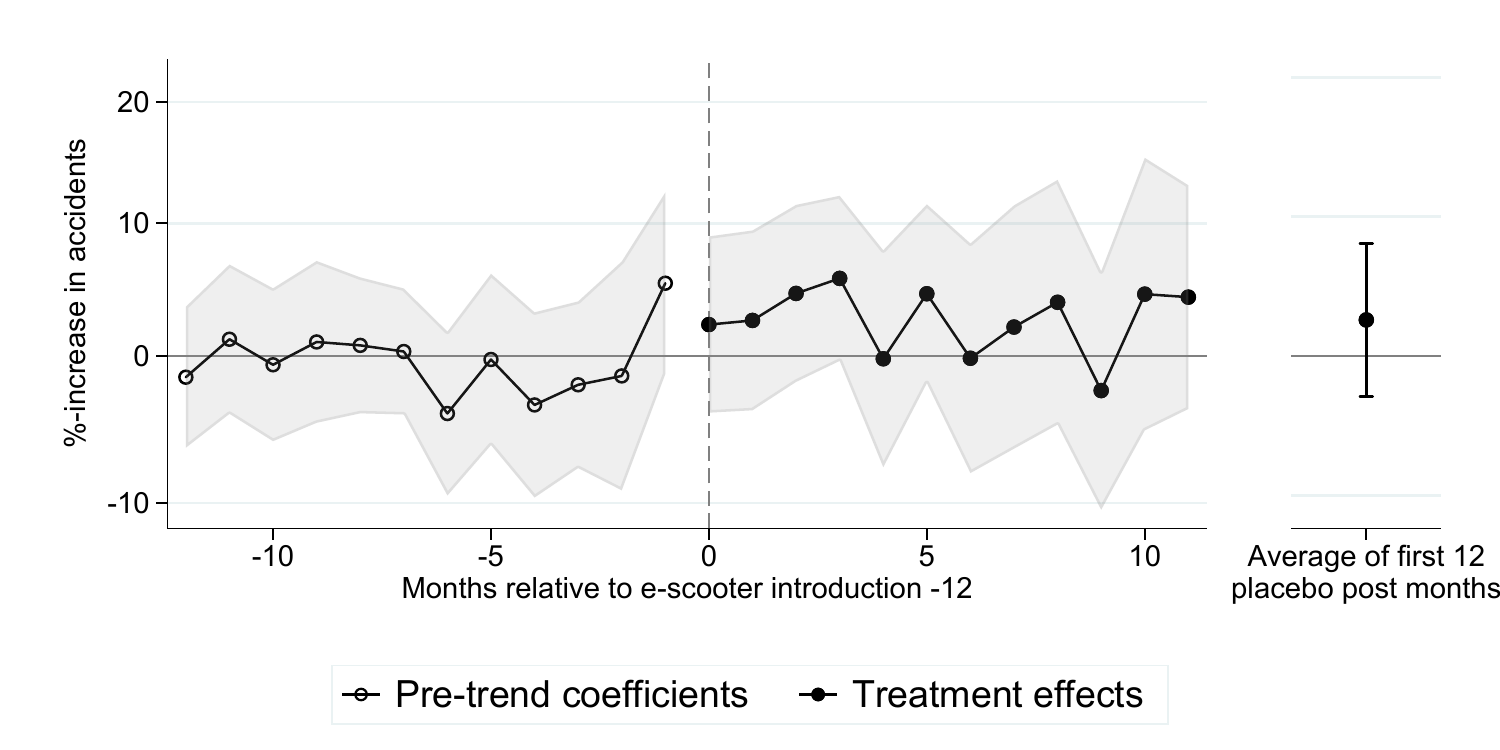}
    \includegraphics[width=0.8\linewidth]{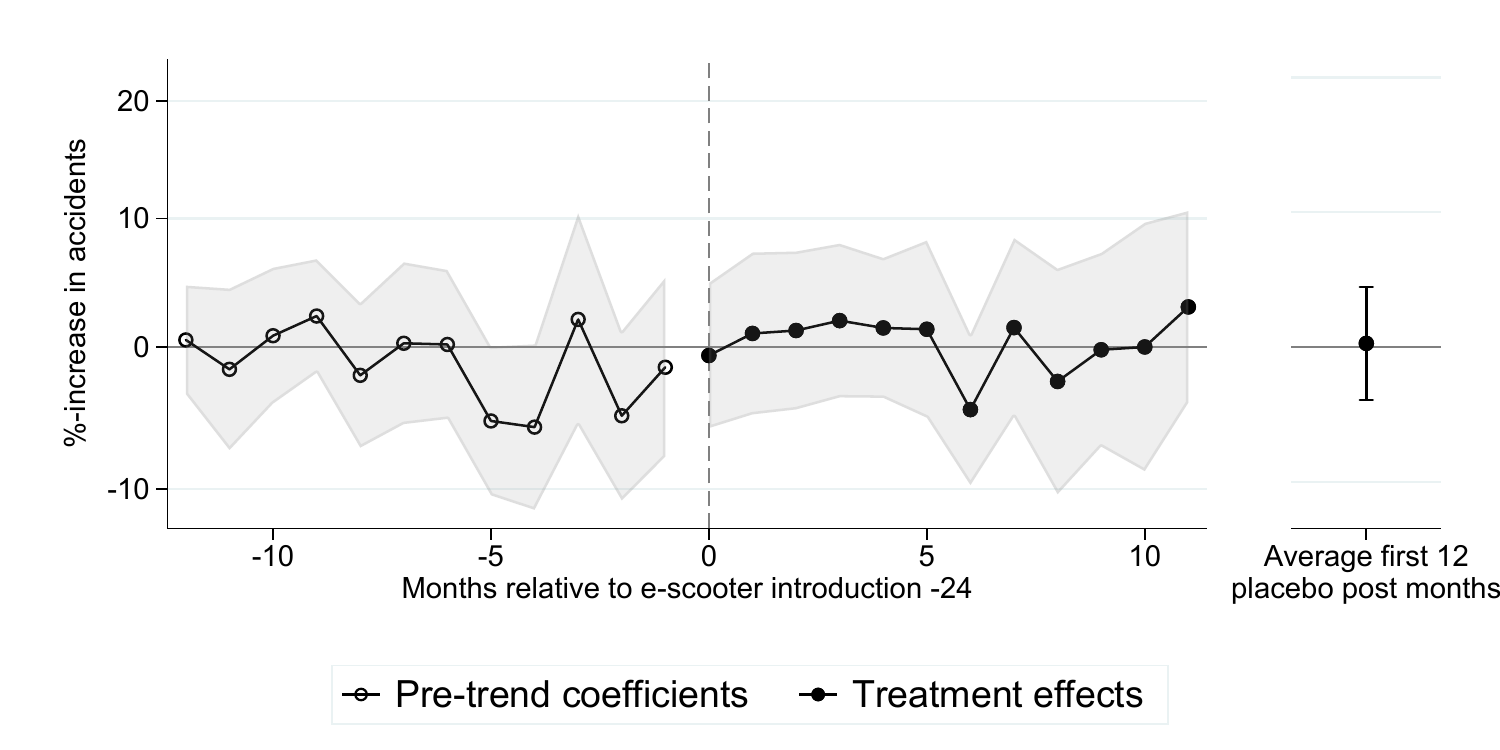}
    \end{center}    
    \label{fig:shiftplacebo}
        \begin{tablenotes}[flushleft]\footnotesize
        \item \emph{Notes:} In each left panel, filled dots indicate treatment effect estimates by month since the placebo-introduction of e-scooter services. Circles indicate estimates for pre-placebo-introduction differences. Shaded areas indicate the 95\%-confidence interval around the placobo-estimates, based on leave-out standard errors defining half-year of treatment launch as cohorts, as described in \cref{section:methods}.
        In the right panels, the average placebo-treatment effect estimates across all cities and the first 12 available post months are indicated (analogous to \cref{tab:22did}, column 2). 
        \end{tablenotes}
\end{figure}

\Cref{fig:shiftplacebo} shows the same analysis as \cref{fig:borusyakplot} but instead of using the actual month of e-scooter introduction, it uses placebo introduction dates that are shifted into the past by 12 or 24 months. Unlike in the real specification, we observe no significant treatment effects in this placebo specification.

\begin{figure}[htbp]
    \caption{After applying the natural logarithm, the seasonality in accident numbers runs parallel.}
    \begin{center}
        \includegraphics[width=0.8\linewidth]{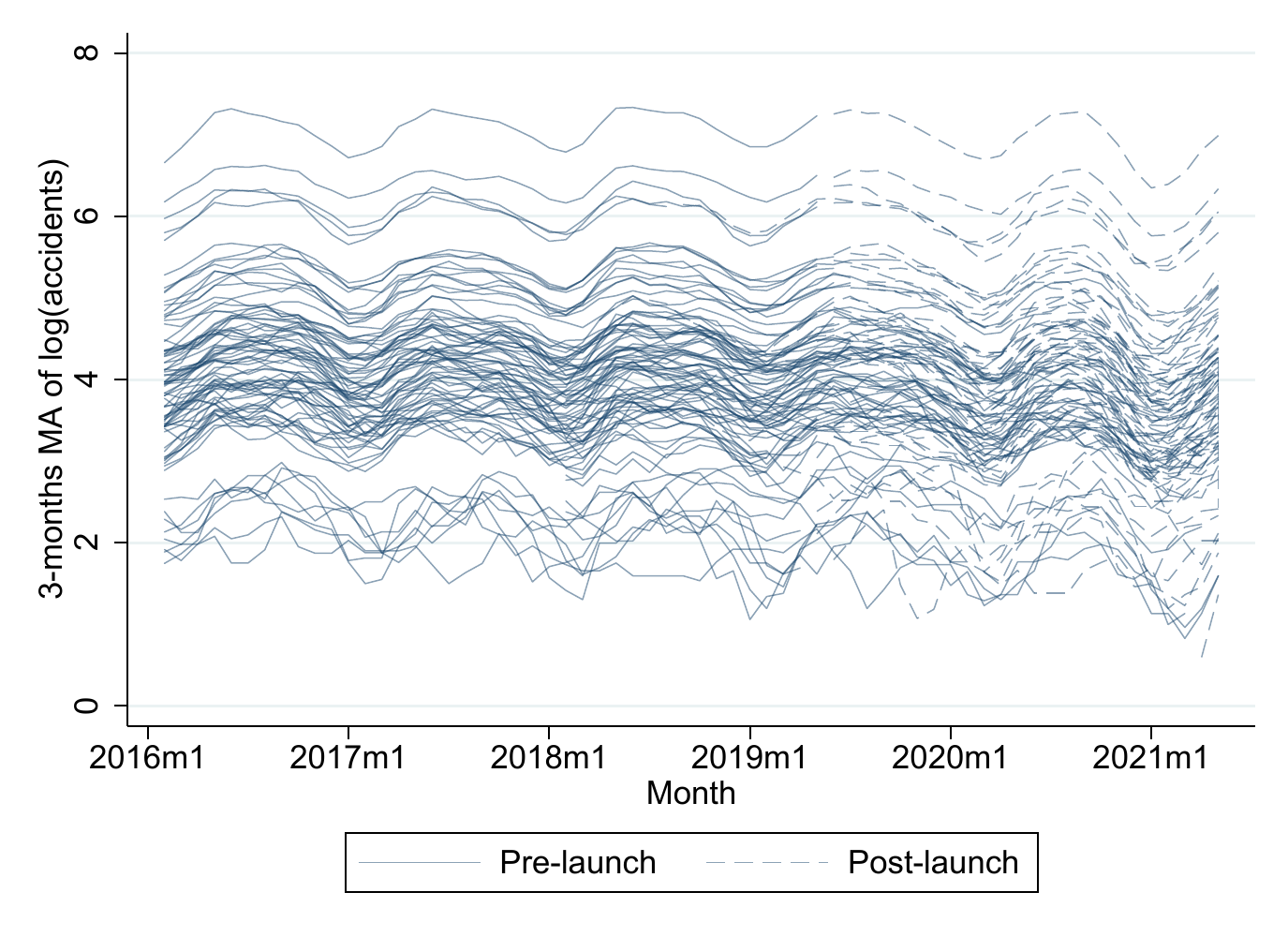}
    \end{center}
    \label{fig:logsparralel}
        \begin{tablenotes}[flushleft]\footnotesize
        \item \emph{Notes:} Each line represents the 3-month moving average of the natural logarithm of accident numbers for one city. Measures in the lower ranges are noisier due to the smaller absolute numbers. Dashed lines indicate accident numbers after the first launch of e-scooter services. 
    \end{tablenotes}
\end{figure}

\Cref{fig:logsparralel} plots city-level 3-months moving averages for the natural logarithm of accident numbers of all sample cities. The moving average is used to smooth out noise. The resulting figure illustrates that the logarithmized accident numbers follow a pattern, which can be well described through city- and month-level fixed effects.

\subsubsection{Accident severity} \label{sec:app-r1}
\begin{figure}[htbp]
	    \caption{Monthly treatment effects and pre-trends  on the percentage of police-reported accidents involving personal injury involving only `slight injury'.}\label{fig:slight}
        \begin{center}
    		\includegraphics[width=0.7\linewidth,trim=0.4cm 0 0 0.4cm,clip]{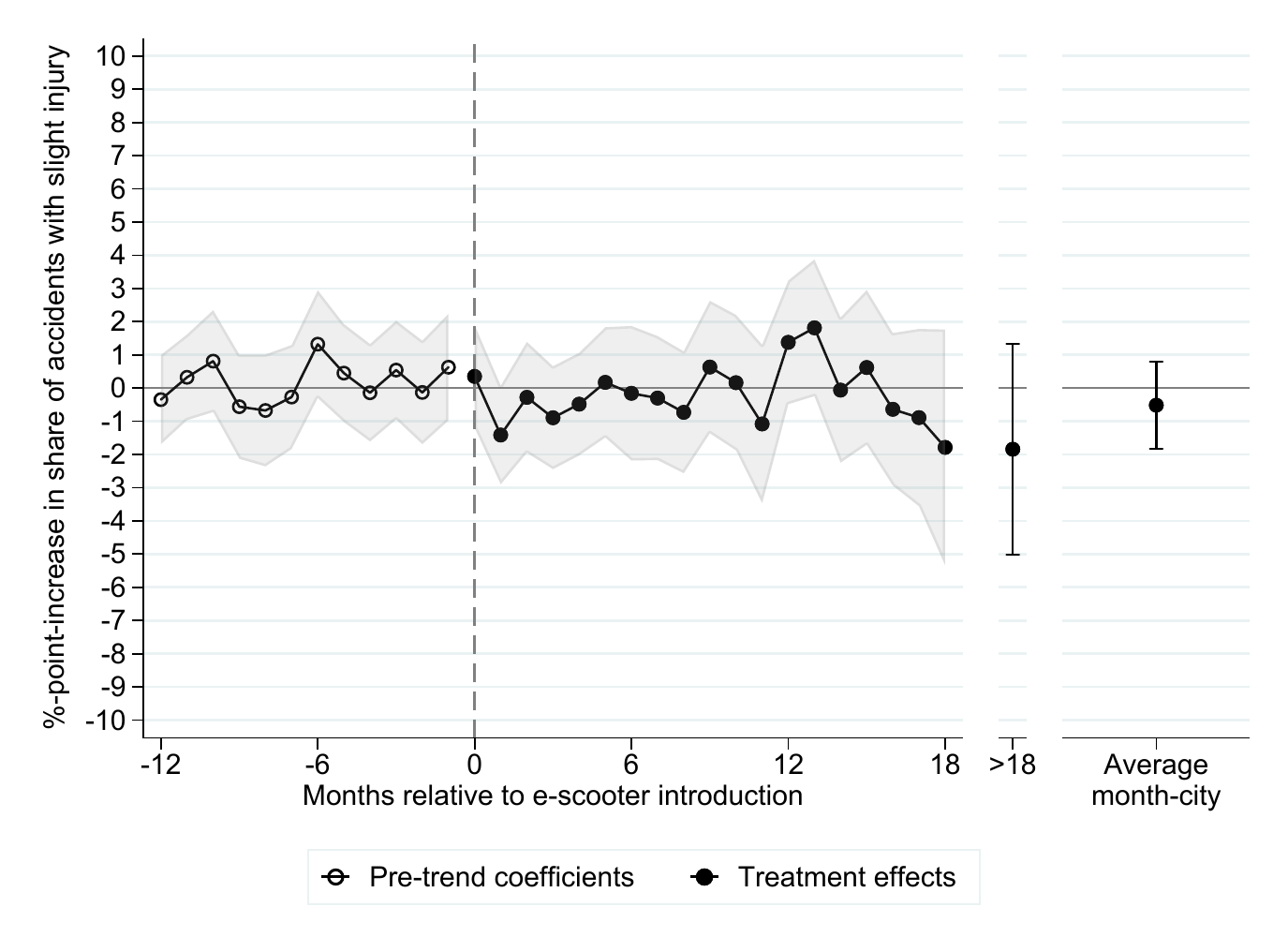}
    	\end{center}
		\begin{tablenotes}[flushleft]\footnotesize
    \item \emph{Notes:}
    The figure shows average treatment effects relative to treatment introduction, on the percentage of accidents with only slight injury. Estimates have to be interpreted as percentage point changes.
    In the line plot, filled dots indicate treatment effect estimates for the first 18 months relative to the introduction of e-scooter services. Circles indicate estimates for pre-treatment months. Effects after month 18 are combined into a single coefficient because month-level estimates for long-term effects are estimated on small subsamples  (few cities had e-scooters early enough for long-term effects to materialize). So, monthly estimates for later months cannot be estimated with comparable precision as for earlier months.
    In the right part, the average treatment effect estimate and corresponding 95\%-confidence interval, across all available post-treatment city-months is shown (\cref{tab:slight}, col.~1). 
    Shaded areas/bars indicate the 95\%-confidence interval around the estimates, based on leave-out standard errors defining half-year of scooter launch as cohorts, as described in materials and methods, sub\cref{section:methods}.
    All estimates account for city fixed effects and month fixed effects.
    \end{tablenotes}
	\end{figure}

We investigate whether there is evidence that the average police-reported accident in a city after the introduction of e-scooters is significantly more or less severe.
If this was the case, this would raise the question if the increase in accident numbers provides an indication for an increase harm. We find no evidence supporting a decrease in severity of reported accidents.
\Cref{fig:slight,tab:slight} show estimates of the treatment effects on the percentage of accidents with personal injuries that involve only `slight injury'. These estimates provide suggestive evidence that, while the total number of accidents changed with the introduction of e-scooters, the composition of the accidents in terms of severity did not change. 

To obtain these treatment effect estimates, a month-city level estimate of the percentage accidents involving slight injury versus severe injury or death is required. To construct estimates for this, we rely on the classifications by the local statistical agencies, as detailed in \cref{section:traffic_data}. While all countries report aggregated accident numbers, as used in the main analysis, countries differ in whether the numbers disaggregated  by severity of injuries refer to the number of injured people or the number accidents. 
Austria, Germany, and Norway report the numbers of victims by severity of injury. Germany additionally reports the number of accidents, for the years covered in the Unfallatlas \citep{unfallatlas} (see \Cref{app-data}). Finland, Sweden and Switzerland only report disaggregated numbers of accidents, i.e., for some countries the data allow us to compute the percentage of accidents with slight injury, while for others we can compute the percentage of victims with slight injury. For the 80 German cities we observe both. In these 80 cities, the two measures have similar means (85.3\% of accidents involve severe injuries vs 87.1\% of victims are severely injured) and a correlation coefficient of 90\%. 
We combine the estimates across all six countries. We use linear regression, estimated on the sample of German observations where we observe the percentage of accidents as the percentage of victims with slight injury to project both into the same scale.

\Cref{fig:slight,tab:slight} reveal that there is no evidence that this is the case and that the rate at which accidents are classified as involving slight injuries only remains stable. The estimate in column 1 of \cref{tab:slight} has a corresponding 95\% confidence interval from \(-1.8\) to \(+0.8\)\%-points), based on which we can reject even moderate decreases in severity, thus see no grounds to assume that the increase in accidents involving e-scooters goes along with a decrease in accident severity. Note also that all our specifications include city-fixed effects that capture potential time-constant differences arising from estimating accident numbers from injury numbers.

    \begin{table}[htbp]
    	\caption{Estimated treatment effects on the percentage of police-reported accidents involving personal injury involving only `slight injury'.}	\label{tab:slight}	
    	\centering
    \begin{center}
    \adjustbox{max width=\linewidth}{\input{tables/22did-r1.tex}}
    \end{center}
        \begin{tablenotes}[flushleft]\footnotesize
    \item \emph{Notes:}
    * p<0.1, ** p<0.05, *** p<0.01. 
    This table shows estimated treatment effects from a specification similar to the one described in materials and methods, \cref{section:methods}, but using a percentage as the dependent variable. The estimation sample restrictions and weights correspond to \cref{tab:22did}, columns 1-3.
    Standard errors are computed using the leave-out procedure recommended in \citet{borusyak2021revisiting}, defining cohorts based on the quarter in which scooters were launched.
    \end{tablenotes}
    \end{table}
	
\subsubsection{Controlling for city-specific time-trends}\label{sec:app-r2}
    \Cref{tab:trends_controls} shows variants of our main treatment effect estimates, controlling for linear time trends in the natural logarithm of accidents. One possible concern is that population (density) growth may affect accidents. If population growth was correlated with the rollout of e-scooters, this could potentially constitute a source for omitted variable bias in our estimation framework. To address this concern, we conducted analyses allowing for city-specific time trends in accident numbers. Population growth is likely be relatively continuous within a time frame of four years, and thus, effects of population numbers on accident rates should be captured by linear city-specific time trends (note that city- and month-specific fixed effects are also still accounted for).

    \begin{table}[htbp]
		    \caption{Replication of main results, while allowing for city-specific time trends.}
    \label{tab:trends_controls}
    \begin{center}    
    \adjustbox{max width=\linewidth}{
					\input{tables/22did-r2.tex}    }
    \end{center}
        \begin{tablenotes}[flushleft]\footnotesize
      \item \emph{Notes:}
    * p<0.1, ** p<0.05, *** p<0.01. 
    This table shows estimated treatment effects from a specifications similar to the log-linear specification described in materials and methods, \cref{section:methods}. However, instead of using the natural logarithm of accident numbers as dependent variable, these estimates use the percentage of accidents with only slight injury. The estimation sample specifications  correspond to \cref{tab:22did}, columns 1-3.
    Standard errors are computed using the leave-out procedure recommended in \citet{borusyak2021revisiting}, defining cohorts based on the quarter in which scooters were launched.
    \end{tablenotes}
\end{table}

%% file: tables/early_or_long.tex
{
\def\sym#1{\ifmmode^{#1}\else\(^{#1}\)\fi}
\begin{tabular}{l*{4}{c}}
\toprule
                    &\multicolumn{1}{c}{(1)}   &\multicolumn{1}{c}{(2)}   &\multicolumn{1}{c}{(3)}   &\multicolumn{1}{c}{(4)}   \\
                    &\shortstack{Months 0--11}   &\shortstack{Months 12--23}   &\shortstack{Early adopters}   &\shortstack{Late adopters}   \\
\midrule
\%-increase in accidents&         5.3** &        10.8***&         4.6** &         6.3** \\
                    &       (2.1)   &       (4.1)   &       (2.3)   &       (3.2)   \\
\midrule
Cities              &          93   &          93   &          93   &          93   \\
Total observations  &        5880   &        5880   &        5880   &        5880   \\
\bottomrule
\end{tabular}
}

%% file: tables/het_did.tex
{
\def\sym#1{\ifmmode^{#1}\else\(^{#1}\)\fi}
\begin{tabular}{l*{3}{c}}
\toprule
                    &\multicolumn{1}{c}{\shortstack{Share of\\bike lanes}}&\multicolumn{1}{c}{\shortstack{Cars per\\capita}}&\multicolumn{1}{c}{\shortstack{Cycling\\modal share}}\\\cmidrule(lr){2-2}\cmidrule(lr){3-3}\cmidrule(lr){4-4}
                    &\multicolumn{1}{c}{(1)}   &\multicolumn{1}{c}{(2)}   &\multicolumn{1}{c}{(3)}   \\
\midrule
Below-median:\\\%-increase in accidents&        11.7** &         2.7   &         9.0***\\
                    &       (4.8)   &       (4.0)   &       (2.9)   \\
\addlinespace
Above-median:\\\%-increase in accidents&         5.3   &        15.2***&         9.7** \\
                    &       (3.8)   &       (4.7)   &       (5.0)   \\
\midrule
\(p\)-value \(H_o\): coefficients identical&        0.29   &        0.05   &        0.91   \\
Cities              &          75   &          75   &          75   \\
Observations        &         150   &         150   &         150   \\
\bottomrule
\end{tabular}
}

%% file: tables/22did-r1.tex
{
\def\sym#1{\ifmmode^{#1}\else\(^{#1}\)\fi}
\begin{tabular}{l*{3}{c}}
\toprule
                    &\multicolumn{1}{c}{(1)}   &\multicolumn{1}{c}{(2)}   &\multicolumn{1}{c}{(3)}   \\
                    &               &\shortstack{Incl. never-\\treated cities}   &\shortstack{First 12\\months}   \\
\midrule
\%-point change     &        -0.5   &         0.4   &        -0.3   \\
                    &       (0.7)   &       (0.5)   &       (0.5)   \\
\midrule
Pre-treat. mean     &\phantom{space}\clap{85.4\%}\phantom{space}   &\phantom{space}\clap{84.9\%}\phantom{space}   &\phantom{space}\clap{85.4\%}\phantom{space}   \\
Treated observations&\phantom{space}\clap{1704}\phantom{space}   &\phantom{space}\clap{1704}\phantom{space}   &\phantom{space}\clap{956}\phantom{space}   \\
Total observations  &\phantom{space}\clap{5879}\phantom{space}   &\phantom{space}\clap{7133}\phantom{space}   &\phantom{space}\clap{5879}\phantom{space}   \\
Cities              &\phantom{space}\clap{93}\phantom{space}   &\phantom{space}\clap{112}\phantom{space}   &\phantom{space}\clap{93}\phantom{space}   \\
\bottomrule
\end{tabular}
}

%% file: tables/22did-r2.tex
{
\def\sym#1{\ifmmode^{#1}\else\(^{#1}\)\fi}
\begin{tabular}{l*{3}{c}}
\toprule
                    &\multicolumn{1}{c}{(1)}   &\multicolumn{1}{c}{(2)}   &\multicolumn{1}{c}{(3)}   \\
                    &               &\shortstack{Incl. never-\\treated cities}   &\shortstack{First 12\\months}   \\
\midrule
\%-increase in accidents&         9.3** &         8.5** &         4.7** \\
                    &       (4.0)   &       (3.5)   &       (2.4)   \\
\midrule
Mean pre-treatment accidents&\phantom{space}\clap{93.2}\phantom{space}   &\phantom{space}\clap{83.0}\phantom{space}   &\phantom{space}\clap{93.2}\phantom{space}   \\
Treated observations&\phantom{space}\clap{1704}\phantom{space}   &\phantom{space}\clap{1704}\phantom{space}   &\phantom{space}\clap{956}\phantom{space}   \\
Total observations  &\phantom{space}\clap{5880}\phantom{space}   &\phantom{space}\clap{7134}\phantom{space}   &\phantom{space}\clap{5880}\phantom{space}   \\
Cities              &\phantom{space}\clap{93}\phantom{space}   &\phantom{space}\clap{112}\phantom{space}   &\phantom{space}\clap{93}\phantom{space}   \\
\bottomrule
\end{tabular}
}

%% file: app-synth-did.tex
\begin{table}
\caption{Synthetic difference-in-differences estimate \citep{arkhangelsky2021synthetic}}\label{table:sdid}

\adjustbox{max width=\linewidth}{
\begin{tabular}{l*{4}{c}}
\toprule
Estimate: &\multicolumn{3}{c}{synthetic difference-in-differences}&\multicolumn{1}{c}{event-study}\\\cmidrule(lr){2-4}\cmidrule(lr){5-5}
\multicolumn{1}{l}{} & (1) & (2) & (3) & (4)\\ 
 & 2016--2020 & \shortstack{2018-2020\\incl.~Sweden} & \shortstack{2016--2020\\excl. winter} & \shortstack{2016--2020, incl.\\Sweden \& Gütersloh}\\ 
\midrule
Treatment effect & 0.041 & 0.0544 & 0.0463   & 0.0564\\ 
 & (0.0173) & (0.0167) & (0.0153) & (0.0213) \\
 \midrule
Monthly MSPE & 0.0168 & 0.0184 & 0.0114  \\ 
\midrule

Treated cities & 71 & 79 & 67 & 80 \\ 
Total cities & 102 & 111 & 102 & 93\\ 
Months in sample & 60 & 36 & 40 & 60 \\
Post-treatment observations & 1022 & 1187 & 680  & 1202\\ 
Full year & Yes & Yes & No & Yes \\ 
\bottomrule
\end{tabular}
}
        \begin{tablenotes}[flushleft]\footnotesize
\item \emph{Notes}: * p<0.1, ** p<0.05, *** p<0.01. 
 Numbers are untransformed coefficient estimates from a log-linear specification. Standard errors in parentheses. 
    Standard errors are computed using the placebo variance estimation method from \citet{arkhangelsky2021synthetic} with 250 replications using \citet{synthdid}.
    Monthly MSPE is the mean squared prediction error of the synthetic control for pre-treatment outcomes, normalized by the number of pre-treatment months. MSPE is included here as a measure of fit of the synthetic control.\\
\end{tablenotes}
\end{table}

Our main specification relies on an assumption of parallel trends for causal identification and we presented several tests as evidence for parallel trends. As an additional robustness check, we show in \cref{table:sdid} that results are similar when using the synthetic difference-in-differences (SDD) estimator for multiple treatment start periods as described in the appendix of Arkhangelsky \emph{et al.} \citet{arkhangelsky2021synthetic}. The SDD relaxes the parallel trends assumption, which states that in absence of the treatment the group of treated and the group of untreated observations would exhibit the same absolute change over time. The SDD instead makes a weaker assumption: that there exists some weighted average of control units that exhibits the same absolute change over time as the treated units would in absence of treatment. Weights are then computed such that the average outcome for the treated units is approximately parallel to the weighted average for control units before the treatment.
In other words, the synthetic control fits the pre-treatment period outcomes of the treated observations as closely as possible while allowing for a constant difference. The assumption is that the synthetic control would continue to closely approximate the counterfactual trend of the treatment group. The coefficient estimates in \cref{table:sdid} are similar to our main results, though not as large, which can be explained by the fact that the SDD estimates have to rely on a slightly different sample, as discussed below. The event-study and SDD estimates are more similar when the former are restricted to the same time period, ending in December 2020, and the same cities (results not shown).

Aside from imposing a slightly less restrictive version of the parallel trends assumption, 
the SDD for staggered rollout designs, as discussed in the appendix of \citet{arkhangelsky2021synthetic}, has more demanding data requirements. It requires a balanced panel (which excludes Sweden for panels including 2016 and 2017) and that control units must be entirely untreated within the panel's time frame. Recall that in our main analysis, all yet-to-be-treated units are used to impute the month- and city-fixed-effects. To maintain a sufficiently large but comparable set of control cities in the SDD analysis, we restrict the analysis to end in December 2020. This implies that all cities that launched scooters in 2021 (see \cref{fig:launch_dates}) can be used to construct the synthetic control. It also implies, however, that we have fewer cities and fewer periods left to compute the treatment effect from.
This required early cutoff is one reason we do not use the SDD as the main specification. By extending the time frame to include more post-treatment periods, cities that become treated fall out of the donor pool of control cities. Losing the yet-to-be-treated control cities reduces the goodness of fit of the synthetic control (measured in the pre-treatment average monthly mean squared prediction error, MSPE). Also, having a smaller donor pool jeopardizes inference by reducing the units that can be used as placebo-treated units for the placebo variance estimation. Similarly, extending the pre-treatment time period forces cities with missing data for earlier time periods to be excluded due to the balanced panel requirement.


Column 1 of \cref{table:sdid} shows the SDD estimate, approximately 4\%, for the monthly average treatment effect of shared e-scooter services on accidents in treated cities until December 2020. This estimate loosely corresponds to column 1 of \cref{tab:22did}, with the difference that the estimand excludes the treatment effects in 2021 and the treatment effect on Swedish cities, for reasons discussed below. Synthetic controls are fitted on pre-treatment data from January 2016 to the date of introduction, which is a minimum of 27 months. Choosing December 2020 as the sample horizon allows for the estimand to include the effect from the summers of 2019 and 2020 when the majority of our study's cities introduced scooters. The donor pool of cities from which the synthetic control is constructed included 31 cities, including 13 yet-to-be-treated cities. The results do not substantively change when choosing an earlier cut-off (e.g., July 2020 has 45 potential controls but the MSPE does not improve and the estimate is similar). However, choosing a later cut-off quickly excludes the yet-to-be-treated control cities. The synthetic control's goodness of fit worsens (i.e. the MSPE doubles) and the treatment effect cannot be reliably estimated.

Column 1 also excludes all Swedish cities and Gütersloh in Germany because of missing data. Column 2 shrinks the panel to start in January 2018 in order to at least include the Swedish cities. This, unfortunately, reduces the pre-treatment observations for Zurich to 3 months, and Basel, Wien, Stockholm, Göteborg, Malmö, and Uppsala all have less than 12 pre-treatment periods.

The last reason we do not use the SDD as the main specification is its susceptibility to bias when there is greater seasonal movement in the outcome variable in treated cities than in any of the control cities (i.e., outside the convex hull of the controls).
We found that the weighted average of the synthetic controls constructed for both columns 1 and 2 under-predicted accidents in the summer (April-September), and over-predicted in the winter (October-March) by the same magnitude, ~3\%. In other words, the average pre-treatment difference between the synthetic control and the treatment group is very close to zero, but the difference is consistently more in winter and less in summer. This volatility could bias the results if the number of summer and winter months with available data varies across cities in the sample. Balancing the seasons in our panel is the final reason we choose December 2020 as a cutoff. For column 1, summer months compromise 50.1\% of the pre-treatment months and 49.6\% of the post-treatment months. Column 2 is 50.2\% pre-treatment and 49.7\% post-treatment summer months.

There is a risk of relying on an estimator with a synthetic control displaying seasonal bias. That is why our preferred specification for the SDD is Column 3 of \cref{table:sdid}, which removes over-compensating seasonality in the synthetic control by dropping the months of November to February from the analysis, where accidents numbers dip low. The goodness of fit increases by a third, and the synthetic control no longer has a detectable seasonal bias. The drawback is that the estimand is no longer an average of all post-treatment months, but just post-treatment months between March and October, which corresponds with Column 3 of \cref{table:sdid}. The estimate is correspondingly larger, likely due to the now restricted set of better weather months in the estimand. 

Column 4 of \cref{table:sdid} restricts our main specification to also end in December 2020. In terms of sample composition Column 2 and 4 are closest, and their estimates are similar. The only two differences in the samples between columns 2 and 4 are that column 4 includes the city of Gütersloh and that it uses 2016--2017 to estimate the city-fixed effects. Both differences are owed to the fact that the SDD estimator requires a balanced panel.  With the SDD estimator, we also replicated results for (i) winter months, (ii) excluding COVID, and (iii) heterogeneity by bike lanes and found they match our main analysis closely (results not shown). But even though synthetic controls have desirable properties, like a relaxed parallel trends assumption, we leave it as a robustness check due to the limitations regarding time coverage and the workarounds needed to overcome potential seasonality bias.

%% file: app-data.tex
\label{app-data}
\subsection{Scooter service data}
\label{section:scooter_data}
\begin{figure*}[htbp]
    \centering
    \caption{Launch dates of first scooter service by city until 12/2021}
    \label{fig:launch_dates}
        \includegraphics[width=0.99\linewidth]{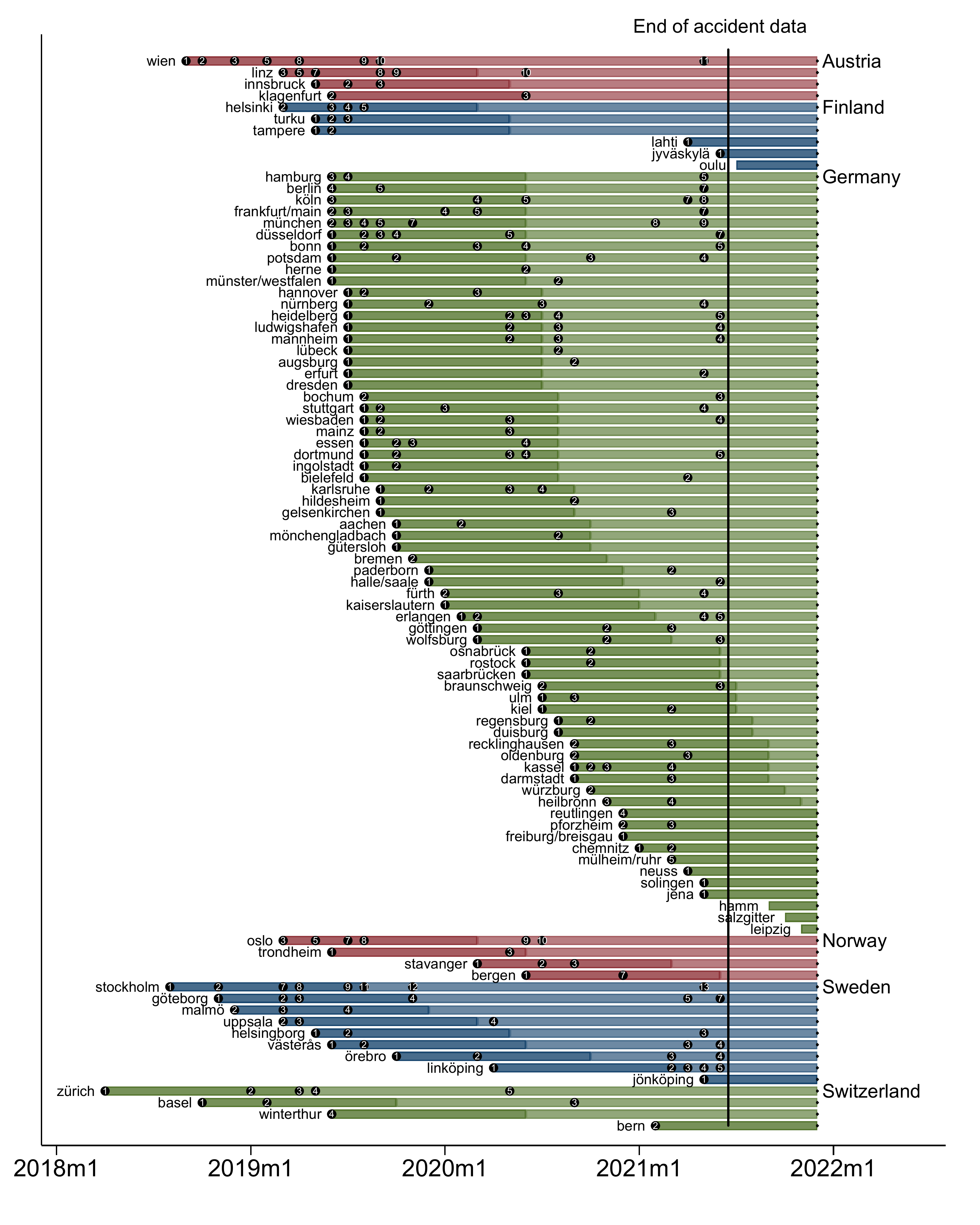}
    \begin{tablenotes}\footnotesize
    \item \emph{Notes}: Bars indicate periods after the introduction of the first scooter service. Black dots indicate the timing of market entry of additional scooter firms until June 2021 (the end of the accident data time frame). Numbers in the black dots indicate the total number of scooter firms that have launched in a city at each point in time.
    Darker shaded areas indicate the first 12 months after the introduction of the first scooter service.
    \end{tablenotes}
\end{figure*}

Figure \ref{fig:launch_dates} provides an overview of the recorded launch dates for different cities in our sample until December 2021. Bars indicate the introduction of the first shared e-scooter service in the respective city, while black dots indicate the timing of additional scooter firms' rollouts. Darker shaded areas illustrate the first year after the rollout of the first shared e-scooter service. Numbers in the black dots show the total number of scooter firms that have launched in a city at each point in time. The vertical black line illustrates the end of our traffic accidents data observation period. The launch dates are hand-collected and verified from official press releases, the firms' social media channels, and websites of e-scooter providers, local newspapers, or municipalities. A file with launch dates for considered providers by city and corresponding sources can be found in the online appendix.

\subsubsection{Contradictions between sources of rollout dates}

Tier and Voi provided us with launch dates for different cities directly. Cities, where the operator Voi stopped operations by 2022, were not included in their list. Furthermore, the dates in provided lists were referring to the latest relaunch dates for cities where services were paused for some time. We included the initial launch date for those cities based on our hand-collected data, after carefully double-checking our sources. The cities are: Linz (Austria), Ingolstadt (Germany), Erfurt (Germany), and Potsdam (Germany). In addition, there are some cases of cities for which the lists provided by Voi and Tier included earlier start dates. These dates seem to refer to the dates when Voi finalized the decision or initiated the launch---not when the first scooters were brought to a city. As we are interested in the timing of the actual rollouts, we carefully cross-checked and looked for additional sources if there were discrepancies. Launch dates may deviate from the provided lists if reliable public sources convincingly confirmed a different effective rollout of e-scooter services in a respective city (more details can be found in the online appendix file). Deviations from the lists provided by Voi and Tier are unlikely to affect our main results, as Tier and Voi were not the first providers in the relevant cities.

Rostock was late to adopt e-scooters. There are, however, some reports of Voi being active in Rostock already in 2019, but according to an event advertised on the provider's and municipality's social media channels, this was just a weekend trial \citep{voi-goes-rostock-e-scooter-pop-up-powered-by-voi}.
We thus use the effective rollout dates according to local newspapers and press releases by the providers (Bird and Moin were the first two providers, with Lime and Tier joining only after the end of the traffic accident observation period).

\subsection{Road accident data}\label{section:traffic_data}

The definition of a road traffic accident with injuries varies between countries in our sample, primarily due to different thresholds for injuries, and inclusion of railed or off-road vehicles in the data. These time-invariant coding differences are absorbed by our model's fixed effects, but to credibly find an effect of shared e-scooters, it is necessary that e-scooter accidents are reported. First, accidents involving e-scooters are eligible to be reported by police, as all countries in our sample define e-scooters in use as moving vehicles. Second, all countries at least record accidents on public roadways, which is inclusive of sidewalks, bike lanes, and medians, as well as car travel lanes, while others include private-areas with public traffic, such as parking lots. Additionally, the sample countries have at a minimum a reporting obligation for police and traffic accident participants for accidents with severe injuries or death, ensuring the vast majority of e-scooter accidents with serious injuries or deaths are reported. However, cycling accidents with single-user and less severe injuries are consistently described as under-reported by the national statistic agencies. The following country sections describe the data generating procedure in greater detail.

We are interested in traffic accidents on the city level. However, data on accidents are reported at different geographic levels of aggregation in different countries. Austria, Germany, and Switzerland's accident data align with city boundaries (see details below). In Finland, Norway, and Sweden, the accident data is reported at the municipality level, which infrequently aligns with city boundaries. Details on countries' traffic accident data and specific cases that were not straightforward are listed and discussed below. We limited our study to cities of at least 100,000 inhabitants. 

A municipality in Scandinavian countries may include multiple localities, fractions of localities, and very remote rural areas. To harmonize the unit of observation across countries, we only consider Scandinavian municipalities that include a locality of at least 100,000 inhabitants. We implement this restriction based on the latest public population statistics for the respective countries. Large urban areas (mainly Helsinki, Stockholm, and Oslo) span multiple periphery municipalities in addition to a core municipality. Because our study is focused on e-scooter service impacts in large cities and to avoid the double-counting of the same urban areas, we exclude ``suburban'' municipalities, which differ in terms of traffic movements and infrastructure. The following municipalities met the population-based exclusion restriction but fall into our definition of suburbs and thus were excluded from our analyses: Norway: Bærum (a suburb of Oslo). Sweden: Huddinge and Nacka (suburbs of Stockholm). Finland: Espoo and Vantaa (suburbs of Helsinki).

\subsubsection{Austria}

Aggregated data on road traffic accidents is published by Statistics Austria on a quarterly basis. 
Statistics Austria provided us with the number of monthly accidents involving personal injuries, segmented by ``politische Bezirke'' (political districts), which is equivalent to the respective (single) municipality for Austrian cities above 100,000 inhabitants. In 2020, a more detailed table section was published \citep{statistikaustria2020}. Our data is equivalently structured as Table 127 of their publication from 2020 \citep{statistikaustria2020}. 
According to the definition by Statistics Austria \citep{stataustria2021doku}, a road traffic accident involving personal injury occurs when one or more persons are injured or killed as a result of traffic on public roads, where at least one moving vehicle was involved. Austria collects accident data on moving vehicles (means of transport for use on roads, including cars, bicycles, and e-scooters) and vehicle-like means of transport (roller-blades, and skateboards). By law, all participants or witnesses of a road-traffic accident must report injuries or deaths to the police, who issue an electronic traffic accident report received by Statistics Austria with a high degree of completeness. Serious injuries result ``in an inability to work or health
problems for more than 24 days''. Injuries of less severity are classified as slightly injured. More information on definitions, comments, methods, and data quality are available from Statistics Austria \citep{stataustria2021doku}.

\subsubsection{Finland}

We obtain data from Statistics Finland which publishes the official statistics on road traffic accidents involving personal injury by municipalities and month. Data is published monthly as preliminary data and  annually as final data \citep{finlandstat}. An accident involving personal injuries is defined by an event  ``that has taken place in an area intended for public transport or generally used for transport and in which at least one of the involved parties has been a moving vehicle'' resulting in death or  personal injury ``which require medical care or observation in hospital, treatment at home (sick leave) or surgical treatment, such as stitches.'' Moving vehicles include micromobility transport, like bicycles and e-scooters. Serious injuries require medical care and ``are classified as serious in accordance with the AIS Abbreviated Injury Scale''. The authors define slight injuries as the remainder of non-serious personal injury accidents. Data is reported by the police through the police information system to Statistics Finland, and data for each month are updated three months after its ending. Deaths are reported with near 100\% accuracy, while accidents with injuries is around 30\%. The worst coverage is for cyclist injured in single road user accidents. Detailed documentation of the data is available \citep{finlandstat2}.

Accidents are reported at the municipality level.
To make the data comparable to other countries, we further make use of the definition of urban settlements provided by Statistics Finland 
\citep{fingrid}. We only consider municipalities that include an urban settlement with more than 100,000 inhabitants. 
In general, an urban settlement may be split between several municipalities.
In our paper, we, thus, focus on the respective municipalities that align best with e-scooter service areas and the focus of our paper. Accordingly, we exclude the municipalities Espoo and Vantaa, which are part of the urban area of Helsinki. In total, there are six Finnish municipalities that we consider: Helsinki, Tampere, Turku, Oulu, Jyväskylä, and Lahti. The vast majority of the population in these municipalities lives in the largest urban settlement. Lahti has the lowest share with
87.4\% and Helsinki has the highest share with 97.7\% \citep{statfin}.

\subsubsection{Germany}
The road traffic accident statistics in Germany include all accidents resulting from vehicular traffic on public roads (and squares) recorded by the police in which at least one person was personally injured 
\citep{destatis1,destatis2}. Serious injury require immediate hospitalization lasting at least 24 hours. Slight injuries are ``any other person injured in a road crash''. Police are required to compile an report for all accidents they become aware of, however the law only requires the reporting of accidents with death or serious injuries \citep{bundesamt1996fach} ( henceforth Fachserie~8). Accidents with only property damage or only involving walkers are not included. Moving vehicles would include micromobility users such as cyclist, e-scooters, and skateboarders. The Federal Statistical Office of Germany publishes a monthly list of accidents in 100 cities (Fachserie~8). That list is unchanged since 2008 and contains all cities above 100,000 inhabitants (except Gütersloh) and a non-random subset of smaller cities. There are cases where traffic accident reports were not provided to the statistical agency (by the police) in time, which means that these accidents are not included in the monthly accident statistics.
These accidents are, however, included in the annually published Unfallatlas \citep{unfallatlas}, which covers most German states. For 2020, all of Germany is covered. For 2019, Mecklenburg-Vorpommern is omitted. For 2018, several states are omitted. For our analyses, we rely on the data from Fachserie~8 for all cities where this is available. We replace these numbers with numbers from the Unfallatlas for 9 city-months, where---based on stark discrepancies in trend---we deem the original data unreliable. Accident data for Gütersloh refers to \citep{unfallatlas} and thus only covers the period from 2018 onwards.
\begin{table}[htb]\caption{Observations for which the data from the \emph{Unfallatlas} \citep{unfallatlas} was chosen over the data from \emph{Fachserie~8} \citep{bundesamt1996fach}}\centering
\begin{tabular}{lcccc}
\toprule
      City &   Year&   Month&   \shortstack{Accidents\\(Unfallatlas)}   & \shortstack{Accidents\\(Fachserie~8)}\\\midrule
      Halle/Saale   &2017 &      6  &                    93       &         22 \\
      Halle/Saale   &2017&       7   &                   63      &           3 \\
      Rostock   &2020   &    5        &              36         &        1 \\
      Rostock   &2020  &     6         &             52        &         5 \\
      Rostock   &2020       &7          &            41       &          0 \\
      Rostock   &2020      & 8           &           58      &           1 \\
      Rostock   &2020     &  9            &          51     &            5 \\
      Rostock   &2020    &  10             &         39    &             1 \\
      Wiesbaden  & 2019 &      7            &          84 &               14\\
  \bottomrule
\end{tabular}
\end{table}
Unfortunately, we cannot rely only on the Unfallatlas only, as within-city cross-temporal comparisons are central to our analysis and the Unfallatlas has no data for several German states in 2018.

The city of Herne was omitted in the June 2017 issue of Fachserie~8.
The city of Göttingen was omitted from the December 2016 issue. Neither are available from the Unfallatlas.
In order to be able to retain the observations from Herne and Göttingen in a balanced panel, we impute accident numbers for the missing month by taking the average of the previous and succeeding months.

\subsubsection{Norway}
Monthly accident data at the municipality level is available from Statistics Norway. The data is limited to accidents reported to the police that ``involve at least one vehicle, and that have taken place on public or private roads, streets or places open to general traffic'' \citep{trafikkulykker-med-personskade}. Moving vehicles include small electric motor vehicles, such as e-scooters, and non-motorized vehicles, such as bicycles. Only accidents that resulted in at least one slight injury are counted, where the minimum threshold, slight injury, is defined as ``Minor fractures, scratches, etc. Hospitalization is not required.'' Severely injured is defined as major injuries, but none life-threatening, or worse.
Accidents with injuries are required to be reported to police, but accidents with single-users and less severe injuries are less likely to be reported to the police, especially injuries with cyclists. Police reports are electronically submitted to Statistics Norway. Preliminary accident figures are published monthly, however, numbers are not finalized until May of the following year. Currently, our 2021 numbers are preliminary.

Accidents are reported at the municipal level, which does not always correspond with city boundaries. Data for population of urban areas, available from Statistics Norway, was used for our exclusion criteria. An urban area is determined by the positioning of buildings and the number of inhabitants and is independent of administrative boundaries 
\citep{tettsteders-befolkning-og-areal}.
In total, we consider 4 urban areas/municipalities with more than 100,000 inhabitants: Bergen, Oslo, Stavanger, and Trondheim. The vast majority of the population (at least 97\%) in these municipalities also live in the main urban area. The urban area of Fredrikstad/Sarpsborg has slightly above 100,000 inhabitants but is not included in our data as it spans two different municipalities with less than 100,000 inhabitants each. Furthermore, Bærum is excluded as a suburb of Oslo.

\subsubsection{Sweden}
Road traffic accidents with personal injury are compiled by the Swedish Transport Agency, Transportstyrelsen, based on accidents reported by the police involving death or injury. The police are obliged to report all accidents with physical injuries, and accidents with injuries must be reported to the police by participants in a certain time-frame. 
Injury assessment is determined by the police, usually on-site. ``An injured person who is not seriously injured is slightly injured''. ``A person who has suffered a fracture, crushing injury, laceration, serious cut injury, concussion or internal injury is considered seriously injured'', or if they are expected to be hospitalized. Only accidents involving moving vehicles (such as a car, bicycle, or e-scooter) in a road traffic area are counted. Transportstyrelsen provided the accident data at the monthly municipality level for 2018-2021, but could not provide data for 2016 and 2017 due to staffing constraints. 

We use population data on localities from Statistics Sweden for our exclusion criteria. They define an urban area/locality as a concentration of buildings not separated by more than 200 meters and with at least 200 inhabitants \citep{scb-localities-and-urban-areas}. In total there are ten localities with more than 100,000 inhabitants. Traffic accident data on all municipalities that refer to these cities are included, except Huddinge which is a part of Stockholm according to the definition by Statistics Sweden. Again, the vast majority of these municipalities' population lives in the main locality with shares ranging from 70\% for Linköping to 96\% for Stockholm.

\subsubsection{Switzerland}

Road traffic accident data for Switzerland is published annually by the Bundesamt für Strassen (ASTRA) via the federal geoportal (``Geoportal des Bundes''). We use data from the traffic accident map, which shows all recorded accidents involving personal injury since 2011 geographically according to specific topics, including date and severity of injuries \citep{swissAccMap}. We use the municipality code, provided for each accident, to aggregate the number of accidents. The municipality code uniquely identifies a respective city, and the municipal boundaries correspond closely to city boundaries for cities with over 100,000 people.

According to the official definition, a road traffic accident is defined as an unforeseen event in a public traffic area that results in property damage and/or personal injury and involves at least one vehicle or vehicle-like device, such as a skateboard or scooter. Participants are required to call emergency services if there is an injury. A slightly injured person is defined as ``anyone with a minor injury, such as superficial skin injury
without significant blood loss or a slight restriction of movement'', that can leave the crash site unaided. A serious injury is a National Advisory Committee for Aeronautics code 3 or higher, equating to at least an ``impairment that prevents normal activities (e.g. unconsciousness and open bone fractures)''. For better comparability with other countries, we only consider accidents involving personal injuries. Data is collected by the police in accordance with an accident recording protocol and stored and maintained centrally in the accident recording system of the Federal Roads Office \citep{astra2018doku}.

\subsection{Heterogeneity variables}\label{section:het_data}
\subsubsection{Share of bike lanes}
The ratio of separated bicycle infrastructure to road network length is compiled with data from \textit{OpenStreetMap} obtained in March 2022 \citep{osmextract}. To determine road network length, distance is calculated from the geometries of \textit{OpenStreetMap} objects where the \texttt{highway} key is tagged \texttt{residential, primary, secondary, tertiary} (including respective link categories) or \texttt{motorway, trunk} (link excluded). Similarly, separated bicycle infrastructure includes objects from the road network set where bicycles are separated or have priority, specifically objects with additional tags \texttt{cycleway=lane}, \texttt{cycleway=share\_busway}, \texttt{bicycle=designated} and \texttt{cyclestreet}. In addition, all objects where \texttt{highway} is tagged \texttt{cycleway} or \texttt{livingstreet} are included. Objects with \texttt{highway} tagged \texttt{footway}, \texttt{path}, or \texttt{track} are included if bicycles are explicitly allowed with the additional tags \texttt{bicycle=designated} or \texttt{bicycle=yes}. We restrict and crop \textit{opensreetmap.org} spatial objects \citep{sf} with high-resolution political boundaries based on the unit of analysis of the traffic accident data, i.e., cities for Germany, municipalities for Finland, Sweden, Switzerland, and Norway, and districts for Austria \citep{germanshape, finshape, ausshape, swissshape, norshape, sweshape}.

\subsubsection{Registered cars per capita}
The variable refers to the number of registered cars by city per 1,000 inhabitants in 2018. Data on all cities, with the exception of Austria and Gütersloh (Germany), is available from EUROSTAT. Details on the data can be found on the EUROSTAT website \citep{eurostatdata}.
For Gütersloh, we use administrative data from the federal state of North-Rhine Westphalia that reports the same measure in their 2018 mobility report \citep[][p. 68]{guetersloh}. Data on Austrian cities for 2018 is obtained from Statistic Austria \citep{stataustriaautos}. 

\subsubsection{Cycling modal share}

The cycling modal share is defined by the share of journeys to work by bicycle. Data is obtained from EUROSTAT where available. Most German and all Swiss cities are included in the EUROSTAT data set. We consider cycling modal shares for the year 2018. Details on the data can be found on the EUROSTAT website \citep{eurostatdata}.

Data on Austrian cities are obtained from official municipality websites and refer to the latest traffic reports of the respective city. A list with the respective sources and year of data collection can be found in the online appendix.

For Norwegian cities, we use the cycling modal shares from a study \citep{tennoy2022urban} examining ``Urban structure and sustainable modes' competitiveness in small and medium-sized Norwegian cities''. They provide modal shares for the largest Norwegian cities using data from the Norwegian National Travel Survey collected in 2013/14 and 2017/18. For Swedish cities, we use data from a study \citep{kenworthy2022exploring} reporting the ``percentage of total daily trips by cycling'' for ten Swedish cities. For Finland, the data are obtained from the regional publications of the Finnish National Travel Survey (FNTS). Detailed information on the FNTS is published by the Finnish Transport and Communications Agency \citep{finnishagency}. Cycling modal share data for the city of Jyväskylä is obtained from the official region travel survey in 2019 \citep{jvaskyla} because detailed data for Jyväskylä was not published as part of the FNTS.